\documentclass[journal]{IEEEtran}
\ifCLASSINFOpdf
\else
\fi
 \usepackage[caption=false,font=footnotesize]{subfig}
\usepackage{amsmath,amsfonts}
\usepackage{algorithmic}
\usepackage{algorithm}
\usepackage{tcolorbox}
\usepackage{array}
\usepackage{textcomp}
\usepackage{stfloats}
\usepackage{url}
\usepackage{verbatim}
\usepackage{graphicx}
\usepackage{multirow}
\usepackage{booktabs}
\usepackage{amsthm} 

\usepackage{subcaption}
\usepackage{multirow}
\usepackage{pifont}
\usepackage{mdframed} 
\usepackage[table]{xcolor}

\makeatletter
\let\NAT@parse\undefined
\makeatother
\usepackage{hyperref}  

\newcommand{\tblue}[1]{#1}

\newcommand{\HIMMD}{HIM$_{MD}$}

\begin{document}
%
\title{Influence Strength Estimation in Hyperbolic Space for Social Influence Maximization}
%
%
%

\author{
    Hongliang Qiao, 
    Shanshan Feng, 
    Min Zhou, 
    Xutao Li, 
    Yunming Ye, 
    Fan Li, 
    Shuo Shang, \\
    and Yew-Soon Ong,~\IEEEmembership{Fellow,~IEEE}
    \thanks{This work was supported in part by National Natural Science Foundation of China (No.62572361) and (No. 52405295), in part by Shenzhen Science and Technology Program No. SYSPG20241211173609009 and in part by a grant from the Research Grants Council of the Hong Kong Special Administrative Region, China (Project No. PolyU 25233824 for ECS project funded in 2024/25 Exercise). \textit{(Corresponding author: Shanshan Feng)}}
    \thanks{Hongliang Qiao and Fan Li are with The Hong Kong Polytechnic University, Hong Kong SAR, China. E-mail: hongliang.qiao@connect.polyu.hk; fan-5.li@polyu.edu.hk}
    \thanks{Shanshan~Feng is with the Wuhan University, Wuhan, China. E-mail: victor\_fengss@whu.edu.cn}
    \thanks{Min Zhou is with the Huawei Noah’s Ark Lab, Shenzhen, China. E-mail: zhoumin27@huawei.com}
    \thanks{Xutao~Li and Yunming~Ye are with the Shenzhen Key Laboratory of Internet Information Collaboration, Harbin Institute of Technology, Shenzhen, China. E-mail: lixutao@hit.edu.cn; yeyunming@hit.edu.cn}
    \thanks{Shuo Shang is with the University of Electronic Science and Technology of China, Chengdu, China. E-mail: jedi.shang@gmail.com}
    \thanks{Yew-Soon Ong is with the Agency for Science, Technology and Research (A*STAR), Singapore, and also with Nanyang Technological University, Singapore. E-mail: asysong@ntu.edu.sg}
    \thanks{Manuscript received April 19, 2005; revised August 26, 2015.}
    }

%
%

\markboth{Journal of \LaTeX\ Class Files,~Vol.~14, No.~8, August~2015}%
{Shell \MakeLowercase{\textit{et al.}}: Bare Demo of IEEEtran.cls for IEEE Journals}
%



\maketitle

\begin{abstract}

The Influence Maximization (IM) problem aims to identify a small set of influential users, as seed users, to maximize their influence spread in a social network. 
Recently, graph representation learning approaches have gained wide attention in the IM field for their ability to encode social influence patterns into user representations, which are then used by various strategies to identify target seed users.
While effective, these graph learning-based IM methods face two main limitations. 
First, they fail to model the influence propagation process explicitly, limiting their ability to capture the essential underlying propagation patterns.
Second, they build representations in Euclidean space, which cannot reflect the latent hierarchical structure of social influence distribution. 
As a result, the learned representations are ineffective in supporting seed user selection.
To address these limitations, we propose a novel hyperbolic learning-based IM method, HIM, which leverages hyperbolic representation learning to estimate users' influence strength from social data, particularly historical propagation processes, for solving IM tasks.
Unlike previous approaches, HIM comprises two key components. 
First, Hyperbolic Influence Representation encodes influence spread patterns from both the social network and influence propagation instances into hyperbolic user representations.
When learning from these data sources, the geometric properties of hyperbolic space naturally place highly influential users closer to the space origin, enabling practical estimation of influence strength from the distances of learned representations.
Second, Adaptive Seed Selection introduces a novel scoring mechanism grounded in estimated influence strength.
It leverages the geometric advantages of hyperbolic space to incrementally refine scores using multiple types of hyperbolic distance information, enabling flexible and effective seed user selection.
Extensive experiments on five network datasets demonstrate the superior effectiveness and efficiency of our method under various diffusion models with both known and unknown propagation parameters, highlighting its potential for solving IM problems in large-scale, real-world social networks.
\end{abstract}


%
\IEEEpeerreviewmaketitle

\section{Introduction}

\IEEEPARstart{A}s an important problem in social network analysis, the Influence Maximization (IM)~\cite{kempe2003im, TKDE18_li2018influence_survey, lihui2023_influence_survey, TKDD23_li2023survey} aims to identify a small set of users as the seed users in a social network to achieve maximum influence spread, i.e., the expected number of users would be activated by the seed users in an influence propagation process.
Solving IM problems is of great significance for understanding complex influence propagation patterns of real-world social networks, which is beneficial to many applications, such as viral marketing \cite{AAAI2018_tang, TKDE_teng_2024_multi}, outbreak detection \cite{leskovec2007CELF, AAAI2024_neophytou}, and disinformation monitoring \cite{budak2011limiting, sharma2021network}.

The IM problem has been widely studied for decades. 
Most traditional methods~\cite{kempe2003im, leskovec2007CELF, tang2015IMM, wang2016BKRIS} solve the IM problem assuming that diffusion parameters are known and fixed.
Given a set of seed users, the diffusion process can be simulated repeatedly to estimate their influence spread and thus identify highly influential combinations.
However, in addition to the high computational cost, these methods rely on strong diffusion model assumptions, restricting their applicability in real-world settings.
Instead, recently, machine learning techniques have been explored in IM research to alleviate assumptions on diffusion parameters in a data-driven manner~\cite{TKDD23_li2023survey}, offering improved generalization capabilities.
Notably, \textbf{graph representation learning} has attracted increasing attention in IM research for its ability to simplify network data processing while preserving complex structural features.
Building on this, several methods~\cite{panagopoulos2020IMINFECTOR, kumar2022gnn, zhang2022GCNM, li2022piano, chen2023ToupleGDD, ling2023icml, hevapathige2024_DeepSN, J_liu_2025_gawf} have been proposed and demonstrated effective performance for the IM problem. 
These methods first extract patterns from different forms of social data to build user representations that capture social influence patterns.
Then, various strategies leverage these learned representations to achieve different IM solutions. 
While effective, most graph learning-based IM methods still suffer from two main limitations.

First, \emph{existing graph learning-based methods overlook the intrinsic structure of influence propagation.}
Most of these methods treat the propagation process as a black box, focusing on fitting outcomes rather than understanding the underlying propagation mechanism.
For example, SGNN~\cite{kumar2022gnn} uses a graph neural network (GNN) to learn the influence spread of individual users based on simulated diffusion processes. 
GCNM~\cite{zhang2022GCNM} trains a graph convolutional network (GCN) using pseudo labels derived from node degrees. 
Another method, DeepIM~\cite{ling2023icml}, uses a GNN to learn seed users and final influence spread pairs in an end-to-end framework.
The lack of explicit modeling of the propagation process limits these methods' ability to capture essential patterns that reflect users' influence spread.
In fact, historical propagation processes encode rich signals that are crucial for understanding how influence spreads.
As a result, these methods struggle to capture users' actual influence dynamics, resulting in suboptimal performance for IM tasks.

Second, \emph{Euclidean representations have limited expressive power to reflect social influence patterns in social data}.
Most methods are built in Euclidean spaces, which fail to capture the complex patterns of social data, especially the hidden hierarchical structures in social influence distribution. 
Real-world social networks inherently possess hierarchical structures~\cite{barabasi1999emergence}.
The influence spread of users generally follows a power-law distribution, where only a small number of users hold significant influence~\cite{verbeek2014metric}.
Representations built in Euclidean space fail to capture these inherent characteristics and thus cannot effectively identify influential users for IM.
Moreover, most graph learning-based IM methods~\cite{kumar2022gnn, zhang2022GCNM, ling2023icml, hevapathige2024_DeepSN} suffer from severe scalability issues due to their reliance on the adjacency matrix of the social network, limiting their applicability to large-scale real-world scenarios.

To address these limitations, we explore the IM problem from a new perspective in estimating users' influence strength from social data using hyperbolic geometry.
Intuitively, the influence spread of a user set is fundamentally determined by their influence strength, which reflects each user's significance in influence propagation.
Users with higher influence strength tend to generate greater influence spread.
We estimate influence strength by learning representations from social data, including social networks and influence propagation instances, in hyperbolic space.
This approach offers several advantages for solving IM problems.
First, it is purely data-driven, solving IM in a diffusion model agnostic manner without relying on any assumptions about diffusion models.
In contrast, some existing learning-based methods~\cite{li2022piano, chen2023ToupleGDD} are still implicitly constrained by diffusion models.
Second, by learning from two data sources, it unifies network structure and propagation dynamics to capture social influence patterns, particularly dynamics revealed by influence propagation instances (i.e., \emph{who influenced whom}), which reflect how influence spreads over time.
Last but not least, compared to Euclidean representations, hyperbolic space can preserve the hierarchical patterns inherent in social influence, facilitating practical estimation of influence strength.
Recent studies~\cite{nickel2017poincare, nickel2018_hype_geo, ICML2023_Yang} have shown that hyperbolic representations effectively capture complex hierarchical structures in various data, which Euclidean ones fail to preserve.
In hyperbolic space, points near the origin represent higher hierarchical levels, while those farther away correspond to lower levels but increase exponentially in number~\cite{TPAMI2021_peng_survey}.
This property implicitly aligns with the hierarchical nature of social influence, making hyperbolic space a compelling choice for modeling user representations.
By embedding users into a hyperbolic space guided by social data, we can preserve the influence hierarchy with the geometric structure of the space.
As a result, the spatial information can serve as a proxy for influence strength, enabling effective and efficient identification of target seed users.

Yet, implementing such a framework poses several non-trivial challenges.
First, we need a proper learning strategy to capture user influence characteristics from social influence data, ensuring that the learned representations effectively capture users' potential spread patterns.
Second, after encoding influence patterns into user representations, influence strength should be properly quantified and assessed, facilitating effective seed user selection.
Third, the proposed solution should not only produce high-quality seed sets with significant influence spread but also be efficient enough to be applied to large-scale social networks.

To address above challenges, we develop a novel \underline{\textbf{H}}yperbolic \underline{\textbf{I}}nfluence \underline{\textbf{M}}aximization (HIM) method.
HIM mainly consists of two parts: 
(1) \emph{Hyperbolic Influence Representation} extracts the hierarchical structure information of social influence distribution in the social network and influence propagation instances to construct hyperbolic user representations.
As learning progresses, social data, network embedding properties, and hyperbolic geometry jointly adjust the distribution of user representations.
Users with strong connections move closer together in the embedding space, preserving local structures. 
At the same time, highly influential users move toward the center of the space, indicating their significant role in influence propagation.
Once finished, users' spatial positions in the hyperbolic space enable us to intuitively and effectively approximate their influence strength.
(2) \emph{Adaptive Seed Selection} incrementally selects seed users based on the distance information of learned user representations.
It first assigns each user an initial score based on their distance to the origin, which reflects their estimated influence strength.
It then considers the hyperbolic distances between users to estimate the potential overlap in their influence spread.
Leveraging both distance information, it dynamically adjusts user scores with a sliding window mechanism, enabling the exploration of more candidates capable of generating greater influence spread.

Our main contributions can be summarized as follows:
\begin{itemize}
     \item We explore the IM problem from a novel perspective. Instead of developing sophisticated models, we focus on leveraging the geometry of hyperbolic space to estimate users' influence strength from partially observed propagation history. This allows us to select influential users in a data-driven and interpretable manner. To our knowledge, this is the first work to study hyperbolic representation learning for IM.
     
    \item We propose HIM, a novel diffusion model agnostic method for the IM problem. HIM first encodes social influence patterns from social networks and influence propagation instances into hyperbolic representations, enabling practical estimation of users' influence strength. It then employs an adaptive selection algorithm that updates scores based on the estimated influence strength, allowing for flexible and effective seed user selection.
    
    \item Extensive experiments show that HIM significantly outperforms baseline methods in influence spread across five real-world datasets. Compared to state-of-the-art learning-based approaches, our method not only achieves greater influence spread but also demonstrates superior scalability. The source code and datasets are available at~\url{https://github.com/PlaymakerQ/HIM}.
\end{itemize}

\section{Related work}

\subsection{Influence Maximization}

The IM problem has been studied for decades~\cite{lihui2023_influence_survey, TKDE18_li2018influence_survey}.
Traditional IM methods can be broadly divided into three categories: 
simulation-based~\cite{kempe2003im, leskovec2007CELF, wang2010-simu}, 
heuristic~\cite{chen2009DegreeDiscountIC, chen2010MIA, chen2010LDAG, goyal2011simpath, IPM_xie_2023_efficient}, 
and sampling-based~\cite{tang2015IMM, wang2016BKRIS, NIPS2024_Rui, WWW2024-Chen}.
Sampling-based methods, such as IMM~\cite{tang2015IMM}, achieve solutions close to theoretical guarantees while maintaining computational efficiency. 
However, they rely on fixed diffusion models to obtain approximation guarantees through repeated simulation of the propagation process.
This limits their generalization, making them hard to apply to real-world scenarios.

Recently, numerous studies have applied various learning techniques to solve problems, offering new insights into IM research~\cite{TKDD23_li2023survey}.
Data-driven methods can bypass propagation model dependency and directly extract complex propagation features from influence data~\cite{du2014learn_dif_model, WWW2024-Huang}.
Among these, graph learning and reinforcement learning (RL) are the most popular~\cite{hevapathige2024_DeepSN, tang2024graph, panagopoulos2024learning, feng2024influence, chowdhury2024deep, hevapathige2024_DeepSN, J_wang_2025_dgn}.
For instance, PIANO~\cite{li2022piano} and ToupleGDD~\cite{chen2023ToupleGDD} combine both to solve the IM problem. 
Yet, the learning processes of RL still depend on prior knowledge of diffusion models to define rewards and incur high computational costs.
In fact, graph learning techniques can independently be leveraged to design solutions.
Kumar et al.~\cite{kumar2022gnn} developed a GNN-based regression model that uses the estimated influence spread of each node as the label. 
Similarly, GCNM \cite{zhang2022GCNM}, based on GCN, learns node representations using node degree as labels and selects seed users via the Mahalanobis distance.
Very recently, DeepIM~\cite{ling2023icml} and DeepSN~\cite{hevapathige2024_DeepSN} are two end-to-end models that learns the correlation between a seed set and its corresponding influence spread.
Despite their effectiveness, most methods overlook the influence propagation process and fail to capture intrinsic social influence features, particularly hierarchical structures, resulting in limited performance.
Additionally, scalability remains another significant challenge for existing methods.

\subsection{Hyperbolic Representation Learning}

Recent studies have shown that hyperbolic spaces can effectively model data with complex hierarchical structures~\cite{sarkar2011low, sala2018representation, chami2020trees}.
Learning methods based on hyperbolic geometry have been widely studied.
For instance, some studies \cite{ganea2018hdnn, gulcehre2018hdnn, liu2019hdnn, chami2019hdnn} extended classic deep neural networks into hyperbolic space, demonstrating outstanding performance in graph-related tasks such as node classification~\cite{WWW2023_fu2023hyperbolic} and link prediction~\cite{AAAI2019_wang2019hyperbolic}. 
Besides, other studies~\cite{ACL20_chami2020low, feng2022role} utilized rotation operations in hyperbolic spaces to depict complex relations and enhance the expressive power of representations for various scenarios.
Application of hyperbolic representation learning has achieved great success in various domains, including recommender systems~\cite{yang2022hicf, wang2023hdnr}, information diffusion prediction~\cite{feng2022h-diffu, CIKM23_qiao}, knowledge graph completion~\cite{ACL20_chami2020low, wang2023mixed}, and others~\cite{song2023hisum, WWW2024_GraphHAM}.
To the best of our knowledge, no prior research has investigated the potential of hyperbolic representation learning for the IM. 
We argue that the inherent hierarchical patterns in social data can be captured in hyperbolic space to infer influence spread patterns effectively.

\section{Preliminary}
\subsection{Problem Definition}
\label{sec:problem}
A social network can be defined as a graph $G=(V, E)$, where $V$ is the set of users (nodes) and $E$ is the set of relationships or connections (edges) between them.
Given a seed set $S \subseteq V$, influence propagation is the process where influence spreads from 
$S$ to other users in the social network, with the final spread outcome governed by a diffusion model $\mathcal{M}$.
In IM, we focus on the total number of people influenced at the end of the propagation process.
The IM problem aims to maximize the number of influenced nodes in $G$ by selecting an optimal seed node set $S^{*} \subseteq V$. 
Let $\sigma(\cdot)$ be the spread function representing the total number of users ultimately influenced by a given seed set. Given a social network $G$, a diffusion model $\mathcal{M}$ with unknown diffusion parameters and a positive integer $k$, the IM problem can be defined as:
\begin{equation}
\begin{aligned}
    \text{argmax}_{S^{*} \subseteq V} \sigma(S^{*}), \: \text{ s.t. } |S^{*}| = k.
    \label{eq:im_def}
\end{aligned}
\end{equation}

\subsection{Diffusion Model Assumption}
\label{sec:assume}
In this work, we make a key assumption when applying hyperbolic representation learning to diffusion model agnostic solutions for the IM problem.

\begin{mdframed}[linewidth=0.8pt]
\textbf{Assumption:} The diffusion process is governed by unknown diffusion rules and parameters, and only historical propagation results are observable.
\end{mdframed}
That is, we aim to solve IM by relying solely on observed influence propagation results (i.e., propagation instances among users), without requiring any information about the underlying diffusion mechanism, including its rules or parameters\footnote{
Note that our method does not rely on any diffusion rules during learning or seed selection. Specific diffusion models (e.g., IC or LT) are introduced only at the evaluation stage, following standard practice, to enable quantitative assessment of IM performance.
}, at any stage of method design.
The actual propagation mechanism is highly complex and cannot be accurately characterized. 
Previous studies commonly assume that diffusion parameters can be fully accessed ~\cite{tang2015IMM, guo2020-RIS, li2022piano, chen2023ToupleGDD}, which is impossible to achieve in real-world scenarios.
Here, we address this limitation by assuming that although the diffusion model has unknown parameters, the influence spread of given seed users can be directly estimated from partially observed historical propagation instances.

An \textbf{influence propagation instance} records the entire process of influence spread from a given seed set in the network, capturing all influence activations.
An \textbf{influence activation} between two users can be represented as a directed edge $(u \rightarrow v)$, indicating that user $u$ successfully influences user $v$ in the process.
Accordingly, an influence propagation instance can be defined as a directed graph $G_D^i = (V_D^i, E_D^i)$, where $V_D^i$ represents the set of users and $E_D^i$ represents all influence activations. Note that $V_D^i \subseteq V$.
The $M$ propagation instances can be represented as a graph set $\mathcal{G}_D = \{ G^1_D, G^2_D, \dots, G^M_D \}$. Besides, we name a $G_D^i$ as a propagation graph.

By learning from influence propagation instances, we can solve the IM problem in a diffusion model agnostic manner.
This is practical in reality, for example, product promotion often relies on the propagation history of similar products.

\subsection{Hyperbolic Geometry}
We chose the Lorentz model as the embedding space due to its computational efficacy and stability~\cite{nickel2018_hype_geo}.
An $n$-dimensional Lorentz model $\mathbb{L}^{n}_{\gamma}$ is defined as the Riemannian manifold 
with constant negative curvature $-1/\gamma$, where $\gamma>0$ determines the curvature scale.
When $\gamma = 1$, the Lorentz model can be viewed as a unit hyperboloid model.
Negative curvature endows hyperbolic space with inherent hierarchical geometry. By defining vectors on a hyperbolic manifold rather than allowing arbitrary distributions as in Euclidean space, embeddings naturally reflect hierarchical structures, making hyperbolic geometry well-suited for modeling social influence hierarchies.
Specifically, the vectors in the Lorentz model satisfy $\mathbb{L}^n_{\gamma} = \{\mathbf{x} \in \mathbb{R}^{n+1}: \langle  \mathbf{x}, \mathbf{x} \rangle_{\mathcal{L}} = -\gamma \}$, 
where $\mathbf{x} = (x_0, x_1, \cdots, x_n ) \in \mathbb{R}^{n+1}$ with $x_0 = \sqrt{\gamma+ \sum^{n}_{i=1}x_i^2} > 0$, and the origin of the space is $\mathbf{o}_{\mathcal{L}} = (\sqrt{\gamma}, 0, \cdots, 0)$. 
$\langle  \cdot, \cdot \rangle_{\mathcal{L}}$ represents the Lorentzian scalar product which is defined as:
$\langle  \mathbf{x}, \mathbf{y} \rangle_{\mathcal{L}} = -x_0 \cdot y_0 + \sum^{n}_{i=1}x_i \cdot y_i$.
For any two points $\mathbf{x}, \mathbf{y} \in \mathbb{L}^n_{\gamma}$, there is $\langle  \mathbf{x}, \mathbf{y} \rangle_{\mathcal{L}} \le -\gamma$. The squared Lorentzian distance is defined as:
$ d^2_{\mathcal{L}}(\mathbf{x}, \mathbf{y}) =-2 \gamma-2\left\langle\mathbf{x}, \mathbf{y} \right\rangle_{\mathcal{L}}$.

In the Lorentz model, the rotation operation~\cite{ACL20_chami2020low, feng2022role} is a kind of Lorentz transformations that preserve the Lorentzian scalar product, and therefore acts as an isometry that leaves hyperbolic distances unchanged.
In other words, the rotation operation repositions a Lorentz embedding without affecting its hierarchy.
The rotation operation can be described by a block-diagonal matrix: $\mathbf{Rot}_{\Theta}= \text{diag} \left(\mathbf{R}\left(\theta_{r, 1}\right), \mathbf{R}\left(\theta_{r, 2}\right), \cdots, \mathbf{R}\left(\theta_{r, n / 2}\right) \right)$, where $\Theta = \{ \theta_{r,1}, \cdots, \theta_{r, n/2} \}$ are the rotation parameters. The $2 \times 2$ block $\mathbf{R}(\theta_{r, i})$ is defined as:
\begin{equation}
\footnotesize
\begin{array}{cc} \mathbf{R}(\theta_{r,i}) = \begin{bmatrix}
    \cos(\theta_{r,i}) & -\sin(\theta_{r,i}) \\
    \sin(\theta_{r,i}) & \cos(\theta_{r,i}) \end{bmatrix}
\end{array}.
\end{equation}
The rotation parameters are set to be learnable and are optimized during training in this work.
To simplify, we use $\mathbf{Rot}(\mathbf{x})$ to denote the rotation operation applied to a given vector $\mathbf{x}$.
More details about hyperbolic space can be found in~\cite{TPAMI2021_peng_survey}.

\section{Methodology}
\label{sec:method}
\begin{figure}[!ht]
\centering
    \includegraphics[width=0.70\columnwidth]{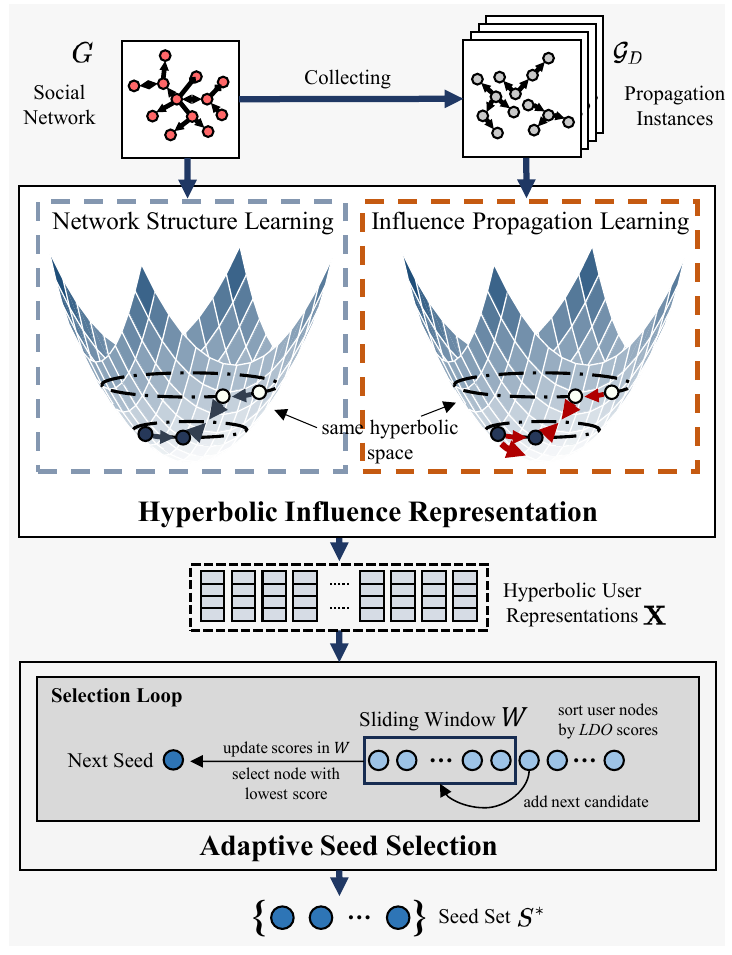}
    \caption{The overall framework of HIM. HIM mainly consists of two modules: (1) \textit{Hyperbolic Influence Representation} aims to learn user representations in the hyperbolic space. (2) \textit{Adaptive Seed Selection} selects target seed users based on learned hyperbolic representations via an adaptive algorithm.}
    \label{fig:HIM}
    \vspace{-2em}
\end{figure}
\subsection{Overview}
We aim to address the IM problem from a new perspective.
We estimate users' influence strength to guide seed set selection.
Specifically, we encode potential influence spread trends into hyperbolic representations for the effective selection of highly influential seed users.
We emphasize two key points of our method:
(1) We develop a fully data-driven, diffusion model agnostic framework for solving the IM problem without any reliance on diffusion parameter assumptions.
(2) The influence spread of users can be efficiently approximated by directly utilizing the properties of learned hyperbolic representations.
We propose HIM, a novel hyperbolic learning method for IM. 
The overall framework of HIM is shown in Figure~\ref{fig:HIM}.

HIM consists of two main components: \emph{learning} and \emph{selection}, which correspond to two modules: Hyperbolic Influence Representation and Adaptive Seed Selection.
In the learning phase, we use the social network and a set of influence propagation graphs as input data, and apply hyperbolic network embedding to construct user representations.
This allows influence strength to be estimated using the distance from the learned representations to the origin of the hyperbolic space.
In the selection phase, we use the estimated influence strength and further consider the propagation relationships between users (measured by hyperbolic distances between their representations) to select a seed set with potential significant influence.
Hence, we can effectively select target seed user nodes using the distance information of the hyperbolic representations.
Instead of explicitly computing users' influence spread through diffusion models, we implicitly estimate it by deriving influence strength from the learned user representations.

\subsection{Hyperbolic Influence Representation}
Hyperbolic Influence Representation encodes social influence features from social influence data to construct user representations in hyperbolic space. 
The social influence data includes social networks and influence propagation instances, as mentioned in Section~\ref{sec:assume}.
Once obtained, these data do not rely on any assumptions about the diffusion model.
That means our learning process is diffusion model agnostic, making it adaptable to various diffusion models and practical applications. 
Both the structural information and the historical spread patterns of propagation instances are crucial for estimating users' influence strength.
We encode these features into the user representations, which enables the estimation of influence strength and, in turn, supports the solution of the IM problem.

The learning process follows a rotation-enhanced hyperbolic network embedding approach.
The properties of hyperbolic network embedding make it particularly well-suited for capturing social influence patterns.
If two users are socially connected, their representations tend to be close to each other in the embedding space. 
Moreover, if a user frequently influences others during the propagation process, their representation tends to move closer to the origin of the hyperbolic space.
We do not adopt more complex embedding methods, such as~\cite{KDD2016_grover_node2vec, KDD2017_ribeiro_struc2vec}, since we aim to intuitively demonstrate that influence strength can be estimated based on hyperbolic representations learned from social influence data, which is previously unexplored.
Meanwhile, this approach maintains computational efficiency, making it scalable for large-scale social networks.
Next, we introduce technical details.

Given social influence data, we propose a rotation-based Lorentz model to learn hyperbolic user representations. 

At first, given a social network $G = (V, E)$, we assign each user $u \in V$ an initial representation $\mathbf{x}_u \in \mathbb{L}^{n}_{\gamma}$, initialized via hyperbolic Gaussian sampling as in work~\cite{sun2021hgcf}.

\subsubsection{Rotation Operation}
We apply the hyperbolic rotation operation~\cite{ICLR19rotate, ACL20_chami2020low} to integrate structural and propagation information for effective representation learning, enabling seamless fusion into unified hyperbolic user representations.

In detail, we use two sets of rotation matrices $(\mathbf{Rot}^{S}_{s}, \mathbf{Rot}^{T}_{s})$ and $(\mathbf{Rot}^{S}_{d}, \mathbf{Rot}^{T}_{d})$ to assist in representation learning. Here, $s$ denotes the social relation, while $d$ denotes the propagation relation. $S$ and $T$ denote the rotation operations applied to head nodes and tail nodes, respectively.
The benefit of rotation operation lies in adjusting vectors' angles to bring related user representations closer while preserving their distances, therefore maintaining hierarchical information.
Besides, it is also efficient and easy to implement.


\subsubsection{Network Structure Learning}
In this part, we deduce structure influence from the social connections present in the social network by modeling the edges within the given graph $G=(V, E)$. 
The core idea is to maximize the joint probability of observing all edges in the graph to learn node embeddings.

Specifically, given an observed edge $(u \rightarrow v) \in E$, the probability $\Pr(v|u)$ can be estimated by a score function based on the squared Lorentzian distance:
$\small \Pr(v|u) = \exp(\mathcal{V}^{S}_{uv})  /  Z(u) $,
where $Z(u) = \sum_{ o \in V } \exp(\mathcal{V}^{S}_{uo})$, and
edge score $\mathcal{V}^{S}_{uv} $ is defined as:
\begin{equation} 
\small \mathcal{V}^{S}_{uv} = - w_{uv} \cdot d^2_{\mathcal{L}}\left(\mathbf{x}^S_u, \mathbf{x}^T_v\right) + b_u + b_v,
\label{eq:relation-score}
\end{equation}
where $ w_{uv} > 0 $ is the coefficient associated with the edge $(u \rightarrow v)$.
Generally, we set $w_{uv} = 1/d_{u}$.
$b_u$ and $b_v$ represent biases of node $u$ and node $v$, respectively.
$\mathbf{x}^S_u = \mathbf{Rot}_{s}^S(\mathbf{x}_u)$ and $\mathbf{x}^T_v = \mathbf{Rot}_{s}^T(\mathbf{x}_v)$ are the rotated representations. 
Since the normalization term $Z(u)$ is expensive to compute, we approximate it via a negative sampling strategy~\cite{mikolov2013neg-sampling}.
Therefore, we estimate $\Pr(v|u)$ in the log form as:
\begin{equation}
\small \log P(v|u) \approx \log \varphi \left(\mathcal{V}_{uv} \right) + \sum_{o \in \mathcal{N}_u} \log \varphi \left( - \mathcal{V}_{uo}\right),
\label{eq:log_p_u_v}
\end{equation}
where $\varphi(x) = 1/(1+e^{-x})$ is the Sigmoid function and $\mathcal{N}_u$ is the set of sampled negative nodes.
Assuming they are independent of each other, the joint probability of all social connections can be calculated as:
\begin{equation} \small \mathcal{P} = \sum_{(u,v)\in E} \log P(v|u). \end{equation}
By maximizing this joint probability, we encode the structure information of the social network into user representations.


\subsubsection{Influence Propagation Learning}
We extract influence spread patterns from the propagation instance graph sets $\mathcal{G}_D$.
Similarly, given any propagation graph $G^i_D \in \mathcal{G}_D$, we maximize the joint probability of observing influence activations in the $G^i_D$ to encode spread patterns into user embeddings.

In detail, given $G^i_D = (V^i_D, E^i_D)$, the edge probability of $(u \rightarrow v) \in E^i_D$ can be calculated similar to Eq. (\ref{eq:log_p_u_v}) as:
\begin{equation}
\small \log P(v|u) \approx \log \varphi \left(\mathcal{V}^{D}_{uv}\right) + \sum_{o \in \mathcal{N}_u} \log \varphi \left( - \mathcal{V}^{D}_{uo}\right),
\end{equation}
\begin{equation}
\small \mathcal{V}^{D}_{uv} = - w_{uv} \cdot d^2_{\mathcal{L}}\left(\mathbf{x}^S_u, \mathbf{x}^T_v\right) + b_u + b_v,
\label{eq:propagation_score}
\end{equation}
where $ w_{uv} = 1/d_u $ is the coefficient, $\mathbf{x}^S_u = \mathbf{Rot}_{d}^S(\mathbf{x}_u)$ and $\mathbf{x}^T_v = \mathbf{Rot}_{d}^T(\mathbf{x}_v)$ are the rotated user representations. 
The joint probability of all edges in $G^{i}_D$ can be calculated as:
\begin{equation} \small \mathcal{P}_{G^i_D} = \sum_{(u,v)\in E^{i}_D} \log P(v|u). \end{equation}

\begin{figure}[ht]
  \centering
  \includegraphics[width=0.8\columnwidth]{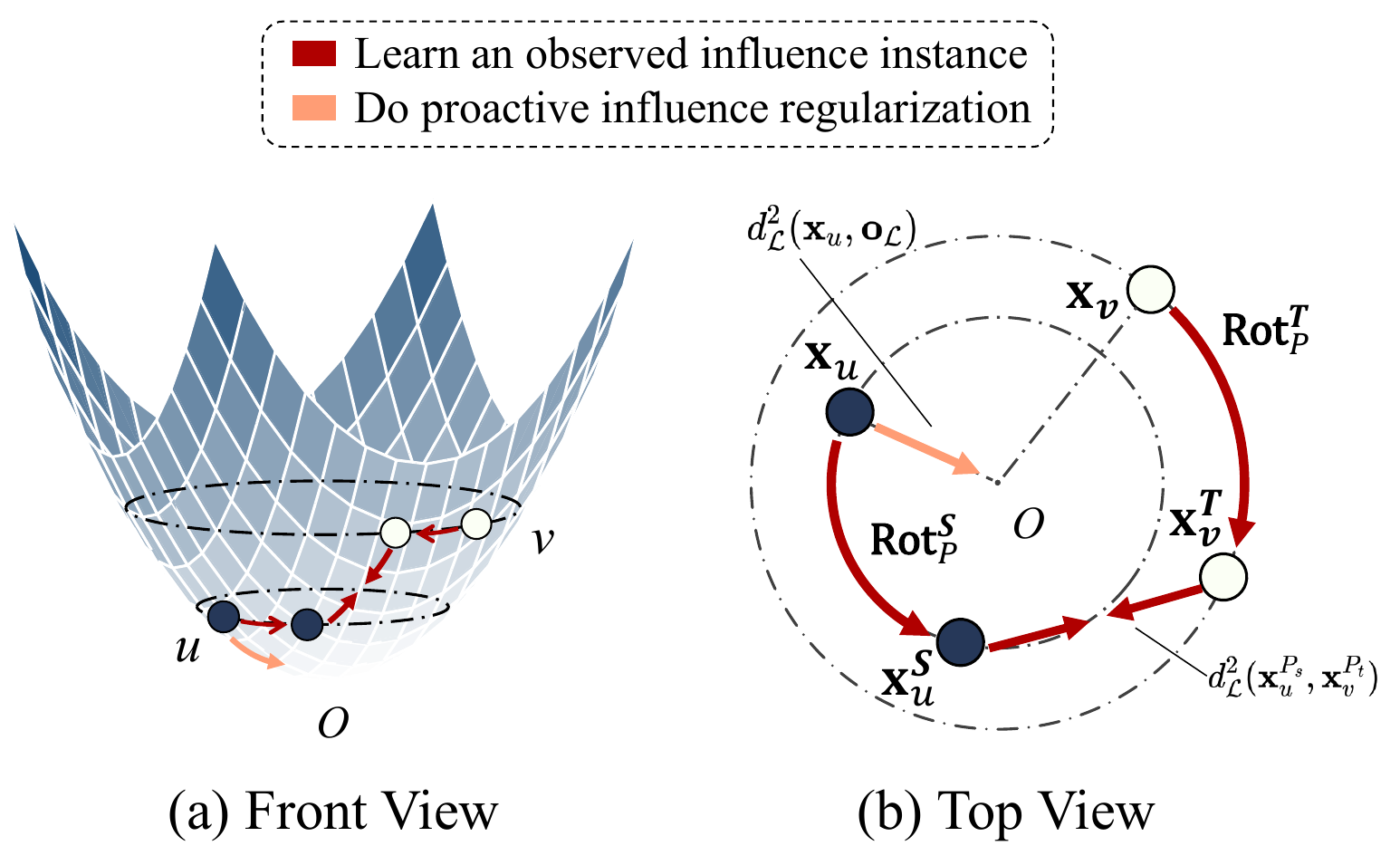}
  \caption{ Illustration of the influence propagation learning. The propagation relation between user $u$ and $v$ is depicted by the distance $d^2_{\mathcal{L}}$ between their rotated embeddings. }
  \label{fig:emb}
\end{figure}

During the propagation process, once a user $u$ triggers influence activation, we want to assign a bonus to highlight this user’s tendency to positively influence others. 
Inspired by the work~\cite{ICML2023_Yang}, we achieve this intuitively by reducing the hyperbolic distance of the related user representations from the origin in the embedding space.
Given influence activations in $G^i_D$, we propose a proactive influence regularization term:
\begin{equation}
 \mathcal{I}_{G^i_D} = \sum_{(u,v)\in G^i_D} \alpha_u \cdot \log \varphi \left( -  d^2_{\mathcal{L}}(\mathbf{x_u}, \mathbf{o}_{\mathcal{L}})\right).
\end{equation}
where $\mathbf{o}_{\mathcal{L}}$ is the origin of the Lorentz model and $\alpha_u$ is calculated as $\sqrt{d_u/d_{\text{max}}}$.
This term further pulls high-influence users closer to the origin in the representation space.
The illustration of learning an observed influence instance $(u \rightarrow v)$ is shown in Figure~\ref{fig:emb}.

For simplicity, we define $LDO$ as the squared Lorentzian distance from a given representation to the origin. Specifically, for user $u$, the $LDO_u$ is defined as $LDO_u = d^2_{\mathcal{L}}(\mathbf{x}_u, \mathbf{o}_{\mathcal{L}})$.
Previous studies~\cite{nickel2017poincare, ICML2023_Yang, feng2022role} have shown that hierarchical information can be effectively inferred from $LDO$s. In our method, user nodes with smaller $LDO$ values are more likely to be influential in social networks. 
We will later design seed selection strategies based on $LDO$.


\subsubsection{Objective Function}

Combining the above two parts, the overall loss function is calculated as:
\begin{equation}
\small
\mathcal{L} = - \left( \mathcal{P} + \sum_{G^i_D \in \mathcal{G}_D}\left(\mathcal{P}_{G^{i}_D} + \mathcal{I}_{G^i_D}\right)\right). 
\label{eq:over_loss}
\end{equation}
Optimizing Eq. (\ref{eq:over_loss}) brings relevant nodes closer together while keeping irrelevant nodes as far apart as possible. Meanwhile, users involved in more influence activations tend to have their representations move closer to the origin, indicating potential higher influence spread.
The learning time complexity is $O( (|E|+|E_D|) \cdot n \cdot |\mathcal{N}| \cdot I )$, where $|E|$ is the number of edges in $G$, $|E_D| = \sum_{i=1}^{M}|E^i_D|$ is the number of all propagation edges,
$n$ is the number of dimensions, 
$\mathcal{N}$ is the number of negative samples,
and $I$ denotes the number of training epochs.

Once the learning process is complete, users with strong spread relations will be clustered together in the embedding space, and highly influential users tend to be located near the origin, which helps to identify seed users for the IM problem.

\subsection{Adaptive Seed Selection} 
\begin{algorithm}[H]
\small
\caption{Adaptive Sliding Window (ASW)}
\label{alg:ASW}
\begin{algorithmic}[1]
\STATE \textbf{Input:} social graph $G$, user representations $\mathbf{X}$, seed number $k$, window coefficient $\beta$ 
\STATE \textbf{Output:} Seed set $S^*$ with $k$ users
\STATE $S^* \leftarrow \emptyset$, window size $w \leftarrow \beta \cdot k$
\STATE $D \leftarrow$ compute $LDO_u = d^2_{\mathcal{L}}(\mathbf{x}_u, \mathbf{o}_{\mathcal{L}})$ for each $u \in V$
\STATE $\mathcal{Z} \leftarrow$ sort $D$ in ascending order
\STATE $c \leftarrow \arg\min_{u} \mathcal{Z}_u$
\STATE $Q \leftarrow$ priority queue with $(u, \mathcal{Z}_u)$ for next $w$ users in $\mathcal{Z}$
\WHILE{$|S^*| < k$}
    \STATE Add $c$ to $S^*$; find $N_c \leftarrow$ neighbors of $c$ in $G$
    \STATE $\mathcal{C} \leftarrow N_c \cap \text{keys}(Q)$
    \IF{$\mathcal{C} = \emptyset$}
        \STATE $c \leftarrow Q.\text{pop}()$
        \STATE Add next $(u, \mathcal{Z}_u)$ from $\mathcal{Z}$ to $Q$
    \ELSE
        \STATE Compute $\mathcal{Z}'_v$ using Eq.~(\ref{eq:update_score}) for each $v \in \mathcal{C}$
        \STATE Update $Q$ with $(v, \mathcal{Z}'_v)$
        \STATE $c \leftarrow Q.\text{pop}()$
        \STATE Add next $(u, \mathcal{Z}_u)$ from $\mathcal{Z}$ to $Q$
    \ENDIF
\ENDWHILE
\STATE \textbf{return} $S^*$
\end{algorithmic}
\end{algorithm}
\vspace{-1em}
After integrating social influence patterns into the hyperbolic representations, the next step involves designing strategies using these representations to select target seed users. 
Given our representation learning objective, an intuitive strategy is to select the 
$k$ users with the lowest $LDO$s.
However, directly selecting users with the highest estimated influence strength may not lead to optimal overall performance due to the submodular nature of influence spread~\cite{kempe2003im}. 
To this end, we propose adaptive seed selection (ASW), which aims to leverage the geometric properties of hyperbolic representations to identify seed users who possess large influence spread effectively.
Specifically, ASW is a lightweight heuristic that explores potential redundancy among seed users by leveraging local hyperbolic distance information, striking a balance between selection effectiveness and computational efficiency.
ASW represents a concrete attempt to leverage the geometric properties of hyperbolic representations for seed selection, and more advanced strategies can be developed by further exploiting this geometric information.

In practice, highly influential users may have overlapping influence spread areas. 
The submodularity property of the IM problem implies diminishing marginal gains from seed users~\cite{TKDE18_li2018influence_survey}, particularly for users who are close to the already selected seed users. 
Therefore, it is crucial to consider these spread relations to measure the proximity between selected seed users. 
Previous methods required traversing all nodes, leading to high computational costs. 
Given that influence strength is estimated by a user representation’s distance from the origin, the distance between various representations can be used to measure spread relations among users.
Based on this insight, we design a new algorithm to refine influence strength estimation for seed set selection by updating an influence score, as shown in Algorithm~\ref{alg:ASW}.

\tblue{The key idea of ASW is to assign each user an initial score and dynamically adjust these scores using the learned hyperbolic distance information between user nodes during the selection process.
Since $LDO$s provide a meaningful ranking of users by influence strength, we restrict exploration to an expanded set of high-potential candidates beyond the top-$k$ nodes, focusing on identifying effective seed combinations within this region.
To this end, we introduce a sliding window that maintains a compact set of top-ranked candidates and dynamically re-evaluates them as new nodes enter.
This enables a more refined re-ranking of candidates based on updated scores while maintaining computational efficiency.}

Specifically, we first assign each user with a score $\mathcal{Z}_u = LDO_u = d^2_{\mathcal{L}}(\mathbf{x}_u, \mathbf{o}_{\mathcal{L}})$. 
We sort all scores $\mathcal{Z}$ and select the node $c$ with the smallest $\mathcal{Z}_c$ as the first seed user. 
Instead of directly choosing the node with the second lowest $LDO$, the next $w$ nodes in the sorted list are viewed as candidate nodes, where $w$ is the size of a sliding window $W$. 
The $W$ is used to explore a broader range of candidate nodes while maintaining computational efficiency. 
We determine the window size $w$ based on $k$ as $w = \beta \cdot k$, allowing it to adaptively adjust its size for different data scales.
In Algorithm~\ref{alg:ASW}, the sliding window $W$ is implemented by a priority queue $Q$.
Next, we find the intersection $\mathcal{C}$ of the current seed node's neighbors with the candidate nodes.
Accordingly, we update the scores of the nodes in $\mathcal{C}$. 
For a user $u \in \mathcal{C}$, the updated score $\mathcal{Z}'_u$ is calculated as:
\begin{equation}
\small
\mathcal{Z}'_u = \mathcal{Z}_u + \frac{w_{c,u}}{d_c} \cdot  \mathcal{Z}_{c},
\label{eq:update_score}
\end{equation}
\begin{equation}
\small
w_{c,u} = \frac{
\exp(1/d^2_{\mathcal{L}}(\mathbf{x}_c, \mathbf{x}_u))
}{ \sum_{v \in \mathcal{C}} \exp(1/d^2_{\mathcal{L}}(\mathbf{x}_c, \mathbf{x}_v))}.
\label{eq:update_score_2}
\end{equation}
where, $c$ denotes the recently selected node, $d_c$ is the degree of node $c$, and $w_{c,u}$ means the weight between them. Intuitively, a node closer to node $c$ may have a larger spread overlap with $c$, leading to a larger penalty from node $c$ and thus increasing its score.
\tblue{The term $\mathcal{Z}_c / d_c$ scales the penalty according to the influence strength of node $c$, thereby propagating the impact of the selected seed to its neighboring candidates. Dividing by $d_c$ further adjusts this effect by accounting for the connectivity of node $c$, which helps to avoid excessively large updates and promotes balanced score updates in practice.}
In this way, the candidates' scores in the sliding window will be updated. After that, we select the node with the lowest score. At each iteration, the chosen node is removed from the window, and the next node from the sorted $LDO$ list is added to the window. This process is repeated until $k$ seed users are selected. 

The time complexity of ASW consists of two main parts. 
Outside the loop, the main time cost is sorting $LDO$ for all nodes, with complexity $O(|V| \log |V|)$. Within the loop, finding neighbors and computing intersections each have a complexity of $O(|N_c|)$, and priority queue updates are $O( |\mathcal{C}| \log w)$. Consequently, ASW's time complexity is $O(|V| \log |V| + k \cdot (2|N_c|+ |\mathcal{C}| \log w))$.

\begin{table}[!tb]
\centering
\vspace{-1em}
\caption{Statistics of datasets.}
\vspace{-1em}
\resizebox{\columnwidth}{!} {
\begin{tabular}{ccccccc}
\toprule
    & Cora-ML 
    & Power Grid 
    & Facebook 
    & GitHub 
    & YouTube
    \\
\midrule
\#Nodes  
& 2,810   
& 4,941    
& 5,908    
& 37,700    
& 1,134,890  
\\
\#Edges  
& 7,981    
& 6,594    
& 41,729   
& 289,003   
& 2,987,624  
\\
$d_{\text{avg}}$ 
& 5.7     
& 2.7      
& 14.1     
& 15.3      
& 5.3        
\\
\bottomrule
\end{tabular} }
\vspace{-2em}
\label{table:datasets}
\end{table}

\section{Experiments}
\label{sec:experiments}
In this section, we conduct extensive experiments to evaluate whether our method can effectively and efficiently identify the most influential seed users across various IM scenarios, including both known and unknown diffusion parameters.
\subsection{Experimental setups}
\subsubsection{Dataset}
We validate our method on five networks of varying sizes, including:
\textbf{Cora-ML}, and \textbf{Power Grid}~\cite{ling2023icml}, 
\textbf{Facebook}~\cite{AAAI2015_nr_data},
\textbf{GitHub}~\cite{rozemberczki2019multiscale},
and \textbf{YouTube}~\cite{yang2012defining}.
YouTube is a large-scale network used to evaluate the scalability of our method.
All datasets are real-world networks with diverse structural characteristics.
The statistics are shown in Table \ref{table:datasets}.

\begin{algorithm}
\small
\caption{Influence Instance Sampling}
\label{alg:sample}
\begin{algorithmic}[1]
\STATE \textbf{Input:} Network $G = (V, E)$; diffusion model $\mathcal{M}$; number of instances $M$; seed ratio $k$
\STATE \textbf{Output:} Propagation graph set $\mathcal{G}_D = \{ G_D^1, \dots, G_D^M \}$
\STATE Initialize $\mathcal{G}_D \leftarrow \emptyset$
\FOR{$i = 1$ to $M$}
    \STATE Sample seed set $S_i \subset V$ with $|S_i| = \lfloor k \cdot |V| \rfloor$
    \STATE Simulate diffusion from $S_i$ under $\mathcal{M}$ to collect activations $\{(u \rightarrow v)\}$
    \STATE Construct propagation graph $G_D^i = (V_D^i, E_D^i)$
    \STATE Append $G_D^i$ to $\mathcal{G}_D$
\ENDFOR
\STATE \textbf{return} $\mathcal{G}_D$
\end{algorithmic}
\end{algorithm}
\vspace{-1em}

\subsubsection{Evaluation Protocol}
Our experimental setup mainly follows the protocols used in~\cite{ling2023icml, hevapathige2024_DeepSN}.
We evaluate our method and baselines under two classic diffusion models (IC and LT). 
We further set up two variants (WIC and WLT) to assess the methods under scenarios with unknown diffusion parameters, which have been rarely investigated.
The details of the diffusion models will be presented later.
We use \textbf{spread ratio} $\delta(S^{*})/|V|$ (in percentage \%) as the primary evaluation metric. 
Similar to~\cite{ling2023icml, hevapathige2024_DeepSN}, for each dataset, we set the seed ratio $k$ to determine the number of target seed nodes as a percentage of the total number of nodes of the network, where $k \in [1\%, 5\%, 10\%]$. 

\subsubsection{Baselines}
We compare HIM with various types of state-of-the-art baseline methods:
(1) \textbf{IMM} \cite{tang2015IMM}  
and \textbf{SubSIM} \cite{guo2020-RIS} are two remarkable traditional methods based on RR set and approximation methods.
They were originally designed for settings where sampling and validation use the same diffusion model (i.e., known diffusion parameters).
(2) \textbf{PIANO} \cite{li2022piano} and \textbf{ToupleGDD} \cite{chen2023ToupleGDD} are learning-based methods based on the reinforcement learning associated with graph representation learning.
(3) \textbf{IMINFECTOR}~\cite{panagopoulos2020IMINFECTOR}, \textbf{SGNN}~\cite{kumar2022gnn}, \textbf{GCNM}~\cite{zhang2022GCNM}, \textbf{DeepIM}~\cite{ling2023icml},
and \textbf{DeepSN}~\cite{hevapathige2024_DeepSN} are graph learning-based methods and serve as our primary comparison baselines.
In addition, we implement \textbf{HIM$_{MD}$} as a variant of HIM that removes adaptive seed selection and instead selects the top $k$ users with the lowest $LDO$s as the seed set.
Note that HIM$_{MD}$ aligns with our representation learning objective, where highly influential users are embedded closer to the origin.

\subsubsection{Diffusion Models}
We use two classic diffusion models, IC and LT, along with their two variants as the underlying diffusion models. 
We first evaluate our method under the standard IC and LT settings, where the diffusion parameters are known. 
This is consistent with previous standard IM studies.
We further extend the evaluation to their variants, where the diffusion parameters are unknown, to assess the effectiveness of our method in more challenging scenarios.

\emph{Basic Diffusion Models}:
We set the diffusion model parameters following the work~\cite{ling2023icml, hevapathige2024_DeepSN}.
For the \textbf{IC} model, the propagation probability $p_{u,v}$ of edge $(u \rightarrow v)$ is set as $1/d^{\text{in}}_v$.
For the \textbf{LT} model, the threshold is uniformly sampled from the range $[0.5, 0.9]$.
We suggest that readers refer to the paper~\cite{kempe2003im} for a detailed introduction to these diffusion models.

\emph{Unknown Diffusion Parameters}:
We introduce two diffusion model variants: WIC and WLT.
(1) \textbf{WIC} denotes the variant of IC with unknown diffusion parameters. 
In WIC, the propagation probability $p_{u,v}$ for each edge $(u \rightarrow v)$ is uniformly sampled from the range $[0, 0.2]$. 
All IM methods do not know the exact propagation probabilities. Yet, they can utilize the influence instances generated by the diffusion model for seed selection. 
For each experiment with a given seed ratio and dataset, all methods are evaluated on the same WIC instance with identical diffusion parameters.
(2) \textbf{WLT} denotes the variant of LT with unknown diffusion parameters. 
The classic LT model assigns each node a threshold, representing the total influence needed from neighbors to activate. The LT model assumes all neighbors have equal influence on a target node. To better reflect real-world complexity, we refer to work~\cite{kempe2003im} and introduce uncertainty by assigning different influence weights to each neighbor while ensuring their sum remains $1$, obtaining the WLT model.
The threshold of WLT is uniformly sampled from the range $[0.5, 0.9]$.

Note that, in WIC and WLT, randomness is introduced through the random instantiation of diffusion parameters, which are sampled once and then fixed for all subsequent propagation realizations. 
This setting evaluates whether learning-based IM methods can learn stable influence patterns solely from observed propagation instances, without access to the underlying diffusion parameters.

\subsubsection{Influence Propagation Instance}
Our work assumes the diffusion model is unknown, but partial propagation data is available. 
This allows our method to address the IM problem in a diffusion model agnostic setting.

In principle, influence propagation instances can be derived from real-world observations. 
However, real-world data cannot directly verify the performance of IM methods. 
Therefore, we adopt diffusion models to simulate propagation instances for experimental evaluation, which serves only as an evaluation tool and does not contradict the diffusion model agnostic assumption.
We simulate $M=30$ influence propagation instances to serve as real-world propagation data for each specific setting.
For each instance, we randomly select seed users based on the given seed ratio and simulate the diffusion process to generate influence activations.
The pseudocode for the sampling process is provided in Algorithm~\ref{alg:sample}.
In the IC and WIC models, an influence activation refers to a successful propagation event between two users, i.e., $(u \rightarrow v)$ indicates that user $u$ activated user $v$.
In the LT and WLT models, a user activation fact may result in multiple activation instances.
For example, if users $a$ and $b$ jointly activate user $c$, the influence activations are recorded as $(a \rightarrow c)$ and $(b \rightarrow c)$.
All graph learning-based methods are evaluated using the same set of influence propagation instances~\footnote{
Note that different methods exploit these instances in different ways.} generated under the corresponding diffusion model, which serve as observed propagation data to assess their generalization ability.

\begin{table*}[!ht]
\begin{center}
\centering
\caption{The overall performance on four regular networks under the IC model. We report average spread ratios (\%) with standard deviations. The best scores are in \textbf{bold}, and the second-best scores are \underline{underlined}.}
\resizebox{\textwidth}{!} {
\begin{tabular}{c ccc ccc ccc ccc } 
\toprule
\multirow{2}{*}{Method} 
        & \multicolumn{3}{c}{Cora-ML} 
        & \multicolumn{3}{c}{Power Grid} 
        & \multicolumn{3}{c}{Facebook}
        & \multicolumn{3}{c}{GitHub} \\ 
        \cmidrule{2-13}

    & 1\% & 5\% & 10\% 
    & 1\% & 5\% & 10\%
    & 1\% & 5\% & 10\%
    & 1\% & 5\% & 10\%  \\
\midrule  

IMM
& \underline{15.81}$_{\pm2.44}$ & 27.88$_{\pm2.28}$ & 37.10$_{\pm2.38}$ 
& 6.87$_{\pm0.67}$ & \underline{24.34}$_{\pm0.75}$ & \textbf{38.87}$_{\pm1.02}$ 
& \underline{41.98}$_{\pm1.80}$ & \textbf{51.21}$_{\pm1.46}$ & \textbf{57.35}$_{\pm0.63}$ 
& 61.36$_{\pm0.34}$ & \underline{64.29}$_{\pm0.23}$ & \underline{67.32}$_{\pm0.12}$ \\

SubSIM
& 15.14$_{\pm2.01}$ & 27.28$_{\pm1.95}$ & 36.89$_{\pm1.62}$ 
& 6.97$_{\pm0.74}$ & \textbf{24.52}$_{\pm1.15}$ & \underline{38.24}$_{\pm0.87}$ 
& \textbf{42.32}$_{\pm1.78}$ & \underline{50.94}$_{\pm1.21}$ & \underline{57.23}$_{\pm1.02}$ 
& \textbf{61.70}$_{\pm0.29}$ & \textbf{64.42}$_{\pm0.26}$ & \textbf{67.33}$_{\pm0.27}$ \\

\midrule

PIANO
& 11.14$_{\pm2.63}$ & 25.96$_{\pm1.74}$ & 37.71$_{\pm1.36}$ 
& 5.10$_{\pm1.06}$ & 18.37$_{\pm1.21}$ & 34.93$_{\pm0.88}$ 
& 37.04$_{\pm3.35}$ & 46.57$_{\pm1.18}$ & 54.13$_{\pm1.02}$ 
& 60.44$_{\pm0.36}$ & 63.09$_{\pm0.22}$ & 64.98$_{\pm0.31}$ \\

ToupleGDD
& 13.95$_{\pm1.96}$ & \underline{28.59}$_{\pm1.75}$ & \textbf{40.10}$_{\pm1.07}$ 
& 6.78$_{\pm0.67}$ & 22.00$_{\pm0.52}$ & 35.08$_{\pm0.94}$ 
& 38.56$_{\pm2.74}$ & 47.38$_{\pm0.90}$ & 55.38$_{\pm0.59}$ 
& 60.70$_{\pm0.26}$ & 63.43$_{\pm0.27}$ & 66.41$_{\pm0.38}$ \\

\midrule

IMINFECTOR
& 9.32$_{\pm2.47}$ & 24.94$_{\pm2.38}$ & 36.96$_{\pm1.94}$ 
& 4.50$_{\pm0.56}$ & 18.53$_{\pm0.99}$ & 31.87$_{\pm0.92}$ 
& 38.74$_{\pm2.61}$ & 48.67$_{\pm1.21}$ & 54.35$_{\pm1.38}$ 
& 60.47$_{\pm0.41}$ & 62.65$_{\pm0.27}$ & 64.37$_{\pm0.39}$ \\

SGNN
& 12.27$_{\pm2.84}$ & 26.32$_{\pm2.67}$ & 37.62$_{\pm2.15}$ 
& 4.74$_{\pm0.72}$ & 18.75$_{\pm0.83}$ & 32.74$_{\pm0.85}$ 
& 38.97$_{\pm2.49}$ & 48.72$_{\pm1.53}$ & 56.70$_{\pm1.21}$ 
& 60.83$_{\pm0.37}$ & 62.35$_{\pm0.26}$ & 65.26$_{\pm0.27}$ \\

GCNM
& 9.96$_{\pm2.45}$ & 23.63$_{\pm2.38}$ & 30.99$_{\pm1.24}$ 
& 4.73$_{\pm0.65}$ & 17.37$_{\pm0.96}$ & 27.53$_{\pm0.74}$ 
& 36.76$_{\pm3.23}$ & 44.25$_{\pm0.79}$ & 50.01$_{\pm1.95}$ 
& 59.14$_{\pm0.67}$ & 61.15$_{\pm0.29}$ & 63.87$_{\pm0.41}$ \\

DeepIM
& 15.09$_{\pm2.13}$ & 28.29$_{\pm0.81}$ & 38.30$_{\pm1.15}$ 
& 6.77$_{\pm0.73}$ & 22.75$_{\pm0.62}$ & 35.34$_{\pm0.51}$ 
& 38.46$_{\pm1.89}$ & 44.58$_{\pm1.99}$ & 51.02$_{\pm0.93}$ 
& 60.63$_{\pm0.34}$ & 62.60$_{\pm0.20}$ & 65.65$_{\pm0.36}$ \\

DeepSN
& 9.38$_{\pm3.01}$ & 27.84$_{\pm1.27}$ & 37.78$_{\pm1.21}$ 
& 5.34$_{\pm0.77}$ & 20.78$_{\pm0.95}$ & 33.24$_{\pm1.02}$ 
& 39.29$_{\pm1.21}$ & 48.83$_{\pm1.31}$ & 55.91$_{\pm0.97}$ 
& 61.21$_{\pm0.29}$ & 62.21$_{\pm0.27}$ & 66.06$_{\pm0.33}$ \\

\midrule

HIM$_{MD}$
& 15.44$_{\pm2.24}$ & 27.82$_{\pm0.81}$ & 38.47$_{\pm0.95}$ 
& \underline{8.10}$_{\pm0.72}$ & 23.81$_{\pm0.61}$ & 36.23$_{\pm0.91}$ 
& 38.94$_{\pm2.31}$ & 49.23$_{\pm1.47}$ & 55.90$_{\pm1.19}$ 
& 60.85$_{\pm0.53}$ & 62.73$_{\pm0.28}$ & 65.92$_{\pm0.35}$ \\

HIM
& \textbf{15.82}$_{\pm1.56}$ & \textbf{28.70}$_{\pm0.94}$ & \underline{39.16}$_{\pm0.88}$ 
& \textbf{8.44}$_{\pm0.89}$ & 24.22$_{\pm0.89}$ & 37.41$_{\pm0.99}$ 
& 39.36$_{\pm2.27}$ & 49.61$_{\pm1.36}$ & 56.76$_{\pm0.99}$ 
& \underline{61.43}$_{\pm0.29}$ & 63.57$_{\pm0.41}$ & 66.51$_{\pm0.32}$ \\

\bottomrule
\end{tabular}}
\label{table:norm-ic}
\end{center}
\vspace{-1em}
\end{table*}

\begin{table*}[!ht]
\begin{center}
\centering
\caption{The overall performance on four regular networks under the LT model. We report average spread ratios (\%) with standard deviations. The best scores are in \textbf{bold}, and the second-best scores are \underline{underlined}.}
\resizebox{\textwidth}{!} {
\begin{tabular}{c ccc ccc ccc ccc } 
\toprule
\multirow{2}{*}{Method} 
        & \multicolumn{3}{c}{Cora-ML} 
        & \multicolumn{3}{c}{Power Grid} 
        & \multicolumn{3}{c}{Facebook}
        & \multicolumn{3}{c}{GitHub} \\ 
        \cmidrule{2-13}

    & 1\% & 5\% & 10\% 
    & 1\% & 5\% & 10\%
    & 1\% & 5\% & 10\%
    & 1\% & 5\% & 10\%  \\
\midrule  

IMM
& 5.94$_{\pm0.14}$ & 16.74$_{\pm0.21}$ & 30.77$_{\pm0.28}$ 
& 2.55$_{\pm0.02}$ & 11.60$_{\pm0.07}$ & 21.98$_{\pm0.17}$ 
& \textbf{3.38}$_{\pm0.06}$ & 12.54$_{\pm0.10}$ & 24.01$_{\pm0.18}$ 
& \underline{14.61}$_{\pm0.11}$ & 32.33$_{\pm0.14}$ & 72.28$_{\pm6.23}$ \\

SubSIM
& \textbf{6.54}$_{\pm0.05}$ & 17.37$_{\pm0.45}$ & 31.05$_{\pm0.31}$ 
& 2.85$_{\pm0.02}$ & 11.58$_{\pm0.06}$ & 22.32$_{\pm0.14}$ 
& \underline{3.31}$_{\pm0.06}$ & 13.02$_{\pm0.07}$ & 23.39$_{\pm0.27}$ 
& \textbf{14.80}$_{\pm0.06}$ & 32.65$_{\pm0.18}$ & 70.75$_{\pm5.49}$ \\

\midrule  

PIANO
& 1.51$_{\pm0.02}$ & 7.78$_{\pm0.08}$ & 14.36$_{\pm0.09}$ 
& 1.70$_{\pm0.03}$ & 8.61$_{\pm0.08}$ & 16.90$_{\pm0.11}$ 
& 3.16$_{\pm0.07}$ & 10.71$_{\pm0.14}$ & 20.13$_{\pm0.24}$ 
& 12.84$_{\pm0.07}$ & 27.30$_{\pm0.14}$ & 38.83$_{\pm0.26}$ \\

ToupleGDD
& 1.73$_{\pm0.01}$ & 7.89$_{\pm0.11}$ & 14.70$_{\pm0.23}$ 
& 1.85$_{\pm0.02}$ & 9.65$_{\pm0.09}$ & 17.52$_{\pm0.18}$ 
& 3.22$_{\pm0.05}$ & 10.86$_{\pm0.21}$ & 20.26$_{\pm0.21}$ 
& 13.77$_{\pm0.05}$ & 28.29$_{\pm0.20}$ & 41.34$_{\pm0.67}$ \\

\midrule  

IMINFECTOR
& 1.07$_{\pm0.01}$ & 5.41$_{\pm0.02}$ & 13.21$_{\pm0.09}$ 
& 1.28$_{\pm0.00}$ & 6.48$_{\pm0.02}$ & 13.34$_{\pm0.07}$ 
& 1.14$_{\pm0.01}$ & 5.45$_{\pm0.02}$ & 10.91$_{\pm0.02}$ 
& 1.18$_{\pm0.01}$ & 5.53$_{\pm0.01}$ & 11.43$_{\pm0.02}$ \\

SGNN
& 1.21$_{\pm0.01}$ & 5.57$_{\pm0.03}$ & 14.01$_{\pm0.11}$ 
& 1.38$_{\pm0.01}$ & 6.18$_{\pm0.02}$ & 13.01$_{\pm0.08}$ 
& 1.18$_{\pm0.01}$ & 5.56$_{\pm0.02}$ & 11.31$_{\pm0.03}$ 
& 1.13$_{\pm0.01}$ & 5.56$_{\pm0.01}$ & 12.04$_{\pm0.02}$ \\

GCNM
& 1.03$_{\pm0.01}$ & 5.19$_{\pm0.01}$ & 10.22$_{\pm0.02}$ 
& 1.14$_{\pm0.03}$ & 5.83$_{\pm0.05}$ & 11.54$_{\pm0.11}$ 
& 1.13$_{\pm0.01}$ & 5.27$_{\pm0.01}$ & 11.88$_{\pm0.34}$ 
& 1.15$_{\pm0.01}$ & 5.14$_{\pm0.01}$ & 11.05$_{\pm0.02}$ \\

DeepIM
& 5.96$_{\pm0.06}$ & \underline{19.08}$_{\pm0.47}$ & \underline{34.13}$_{\pm0.45}$ 
& 2.41$_{\pm0.06}$ & 12.65$_{\pm0.18}$ & 23.42$_{\pm0.28}$ 
& 3.22$_{\pm0.07}$ & 15.72$_{\pm0.75}$ & 30.11$_{\pm0.56}$ 
& 11.62$_{\pm0.08}$ & 34.86$_{\pm0.36}$ & 73.03$_{\pm0.51}$ \\

DeepSN
& 1.85$_{\pm0.01}$ & 5.93$_{\pm0.03}$ & 13.28$_{\pm0.04}$ 
& 1.52$_{\pm0.01}$ & 6.89$_{\pm0.05}$ & 13.47$_{\pm0.15}$ 
& 1.67$_{\pm0.01}$ & 6.15$_{\pm0.02}$ & 11.54$_{\pm0.04}$ 
& 1.31$_{\pm0.03}$ & 5.66$_{\pm0.01}$ & 11.65$_{\pm0.02}$ \\

\midrule  

HIM$_{MD}$
& 5.27$_{\pm0.01}$ & 18.99$_{\pm0.45}$ & 34.09$_{\pm0.63}$ 
& \underline{3.03}$_{\pm0.01}$ & \underline{15.12}$_{\pm0.64}$ & \underline{33.66}$_{\pm0.37}$ 
& 3.25$_{\pm0.05}$ & \underline{15.78}$_{\pm0.37}$ & \underline{31.51}$_{\pm0.62}$ 
& 13.31$_{\pm0.05}$ & \underline{39.31}$_{\pm0.43}$ & \underline{73.39}$_{\pm0.47}$ \\

HIM
& \underline{5.96}$_{\pm0.19}$ & \textbf{19.21}$_{\pm0.36}$ & \textbf{34.41}$_{\pm0.49}$ 
& \textbf{3.25}$_{\pm0.07}$ & \textbf{15.93}$_{\pm0.38}$ & \textbf{34.16}$_{\pm0.83}$ 
& 3.29$_{\pm0.04}$ & \textbf{15.98}$_{\pm0.65}$ & \textbf{32.90}$_{\pm0.41}$ 
& 13.58$_{\pm0.09}$ & \textbf{39.91}$_{\pm0.46}$ & \textbf{74.68}$_{\pm0.63}$ \\

\bottomrule
\end{tabular}}
\label{table:norm-lt}
\end{center}
\vspace{-2em}
\end{table*}
\subsubsection{Implementation Details}
We adopt the same evaluation scheme as DeepIM~\cite{ling2023icml} and DeepSN~\cite{hevapathige2024_DeepSN}.
The baselines are configured according to either public code repositories or the suggestions provided in the original papers.
For IMM and SubSIM, the $\epsilon$ is set to $0.1$.
For all learning-based models, the default number of dimensions $n$ is set to $64$.
In our methods, the curvature scale $\gamma$ is set to 1.
The number of negative samples per positive sample is set to $5$.
For adaptive seed selection, we set the window coefficient $\beta$ as $1.0$ for Cora-ML, and Power Grid, $\beta$ as $0.5$ for Facebook and GitHub, and $\beta$ as $0.1$ for YouTube.
For all learning-based methods, the learning rate is tuned within the range of $0.0001$ to $0.001$, the maximum number of epochs is set to $200$, and an early stopping strategy is adopted where training stops if the loss does not decrease for 10 consecutive epochs.
For each experiment, similar to DeepIM and DeepSN, 
we evaluate chosen seed users through simulation under a diffusion model, running until the whole process is completed.
We record the average influence spread ratio over 100 rounds.
All code is implemented in Python on an Ubuntu server equipped with a 10-core Intel(R) i9-10900X 3.70GHz CPU and one NVIDIA GTX 4090 GPU.

\subsection{Overall Performance}
We conduct experiments on four datasets across four diffusion models, covering IM scenarios with both known and unknown diffusion parameters, to evaluate whether the proposed method can identify highly influential users for effective influence spread.

\begin{table*}[!ht]
\begin{center}
\vspace{-1em}
\centering
\caption{The overall performance on four regular networks under the WIC model. We report average spread ratios (\%) with standard deviations. The best scores are in \textbf{bold}, and the second-best scores are \underline{underlined}.}
\resizebox{\textwidth}{!} {
\begin{tabular}{c ccc ccc ccc ccc } 
\toprule
\multirow{2}{*}{Method} 
        & \multicolumn{3}{c}{Cora-ML} 
        & \multicolumn{3}{c}{Power Grid} 
        & \multicolumn{3}{c}{Facebook}
        & \multicolumn{3}{c}{GitHub} \\ 
        \cmidrule{2-13}

    & 1\% & 5\% & 10\% 
    & 1\% & 5\% & 10\%
    & 1\% & 5\% & 10\%
    & 1\% & 5\% & 10\%  \\
\midrule

IMM
    & 24.34$_{\pm2.51}$ & 31.64$_{\pm1.38}$ & 36.65$_{\pm1.23}$
    & 2.33$_{\pm0.21}$ & 9.25$_{\pm0.37}$ & 16.86$_{\pm0.31}$ 
    & 16.99$_{\pm4.51}$ & 23.47$_{\pm1.43}$ & 28.72$_{\pm1.76}$ 
    & 27.12$_{\pm0.28}$ & 30.66$_{\pm0.23}$ & 33.22$_{\pm0.16}$ \\

SubSIM
    & 21.89$_{\pm4.51}$ & 31.61$_{\pm2.19}$ & 36.06$_{\pm3.10}$ 
    & 2.85$_{\pm0.31}$ & 9.65$_{\pm0.38}$ & 17.17$_{\pm0.32}$ 
    & 16.67$_{\pm3.06}$ & 23.40$_{\pm1.81}$ & 28.95$_{\pm1.83}$ 
    & 26.68$_{\pm0.13}$ & 30.58$_{\pm0.25}$ & 33.21$_{\pm0.20}$ \\

\midrule

PIANO
    & 25.21$_{\pm1.24}$ & 30.14$_{\pm1.19}$ & 35.78$_{\pm0.74}$ 
    & 2.84$_{\pm0.33}$ & 10.23$_{\pm0.39}$ & 17.43$_{\pm0.21}$ 
    & 18.77$_{\pm0.62}$ & 24.78$_{\pm1.01}$ & 29.07$_{\pm0.37}$ 
    & 26.13$_{\pm0.24}$ & 28.51$_{\pm0.25}$ & 35.01$_{\pm0.20}$ \\
    
ToupleGDD
    & 26.74$_{\pm1.39}$ & 30.02$_{\pm2.72}$ & 36.32$_{\pm0.83}$ 
    & 2.90$_{\pm0.31}$ & 10.74$_{\pm0.44}$ & 17.74$_{\pm0.42}$ 
    & 19.32$_{\pm0.90}$ & 25.64$_{\pm0.65}$ & 30.26$_{\pm0.47}$ 
    & 26.62$_{\pm0.29}$ & 29.27$_{\pm0.24}$ & 34.82$_{\pm0.26}$ \\

\midrule

IMINFECTOR
    & 25.80$_{\pm1.37}$ & 31.66$_{\pm1.07}$ & 37.42$_{\pm0.89}$ 
    & 1.65$_{\pm0.12}$ & 7.96$_{\pm0.41}$ & 15.73$_{\pm0.46}$ 
    & 18.43$_{\pm2.07}$ & 23.76$_{\pm0.77}$ & 27.94$_{\pm0.73}$ 
    & 26.43$_{\pm0.12}$ & 30.76$_{\pm0.28}$ & 34.88$_{\pm0.27}$ \\

SGNN
    & 25.95$_{\pm1.58}$ & 31.17$_{\pm1.20}$ & 36.57$_{\pm1.31}$ 
    & 1.88$_{\pm0.15}$ & 10.39$_{\pm0.34}$ & 17.89$_{\pm0.41}$ 
    & 18.23$_{\pm1.44}$ & 22.62$_{\pm1.34}$ & 27.63$_{\pm1.06}$ 
    & 26.62$_{\pm0.23}$ & 29.57$_{\pm0.16}$ & 35.01$_{\pm0.15}$ \\

GCNM
    & 25.50$_{\pm1.73}$ & 30.81$_{\pm1.35}$ & 34.16$_{\pm1.04}$ 
    & 1.94$_{\pm0.28}$ & 8.43$_{\pm0.35}$ & 15.09$_{\pm0.43}$ 
    & 18.57$_{\pm0.93}$ & 22.84$_{\pm0.75}$ & 28.43$_{\pm0.85}$ 
    & 25.72$_{\pm0.17}$ & 27.57$_{\pm0.32}$ & 34.38$_{\pm0.19}$ \\
    
DeepIM
    & 26.00$_{\pm1.28}$ & 30.90$_{\pm1.26}$ & 36.65$_{\pm0.89}$ 
    & 2.84$_{\pm0.23}$ & 11.19$_{\pm0.41}$ & 18.32$_{\pm0.67}$ 
    & 19.67$_{\pm0.90}$ & 24.77$_{\pm0.95}$ & 28.63$_{\pm0.86}$ 
    & 26.75$_{\pm0.24}$ & 27.47$_{\pm0.23}$ & 29.51$_{\pm0.19}$ \\

DeepSN
    & 23.68$_{\pm4.33}$ & 32.19$_{\pm0.90}$ & 37.54$_{\pm1.24}$ 
    & 2.20$_{\pm0.21}$ & 8.67$_{\pm0.40}$ & 16.51$_{\pm0.77}$ 
    & 19.63$_{\pm1.42}$ & 25.78$_{\pm0.84}$ & 32.42$_{\pm0.76}$ 
    & 27.37$_{\pm0.24}$ & 30.81$_{\pm0.23}$ & 34.81$_{\pm0.21}$ \\
    
\midrule
HIM$_{MD}$
    & \underline{27.22}$_{\pm1.21}$ & \underline{32.98}$_{\pm1.04}$ & \underline{38.19}$_{\pm0.76}$ 
    & \underline{3.37}$_{\pm0.18}$ & \underline{11.28}$_{\pm0.42}$ & \underline{18.66}$_{\pm0.39}$ 
    & \underline{20.45}$_{\pm0.90}$ & \underline{26.02}$_{\pm0.56}$ & \underline{30.16}$_{\pm0.61}$ 
    & \underline{27.69}$_{\pm0.18}$ & \underline{31.41}$_{\pm0.22}$ & \underline{36.12}$_{\pm0.15}$ \\  
HIM
    & \textbf{27.65}$_{\pm1.27}$ & \textbf{33.41}$_{\pm0.94}$ & \textbf{38.87}$_{\pm0.96}$
    & \textbf{3.46}$_{\pm0.27}$ & \textbf{11.45}$_{\pm0.46}$ & \textbf{19.13}$_{\pm0.50}$ 
    & \textbf{21.33}$_{\pm1.08}$ & \textbf{26.99}$_{\pm0.64}$ & \textbf{32.69}$_{\pm0.59}$ 
    & \textbf{27.74}$_{\pm0.21}$ & \textbf{31.60}$_{\pm0.31}$ & \textbf{36.27}$_{\pm0.22}$ \\

\bottomrule
\end{tabular}}
\label{table:result-ic}
\end{center}
\vspace{-1em}
\end{table*}
\begin{table*}[!h]
\begin{center}
\centering
\caption{The overall performance on four regular networks under the WLT model. We report average spread ratios (\%) with standard deviations. The best scores are in \textbf{bold}, and the second-best scores are \underline{underlined}.}
\resizebox{\textwidth}{!} {
\begin{tabular}{c ccc ccc ccc ccc } 
\toprule
\multirow{2}{*}{Method} 
        & \multicolumn{3}{c}{Cora-ML} 
        & \multicolumn{3}{c}{Power Grid} 
        & \multicolumn{3}{c}{Facebook}
        & \multicolumn{3}{c}{GitHub} \\ 
        \cmidrule{2-13}

    & 1\% & 5\% & 10\% 
    & 1\% & 5\% & 10\%
    & 1\% & 5\% & 10\%
    & 1\% & 5\% & 10\%  \\
\midrule  

IMM
    & 8.00$_{\pm0.12}$ & 21.65$_{\pm0.44}$ & 37.83$_{\pm0.53}$
    & \textbf{4.09}$_{\pm0.11}$ & 15.30$_{\pm0.19}$ & 28.58$_{\pm0.33}$ 
    & 4.21$_{\pm0.09}$ & 15.05$_{\pm0.19}$ & 28.78$_{\pm0.28}$ 
    & 17.89$_{\pm0.09}$ & 36.84$_{\pm0.17}$ & 79.34$_{\pm1.64}$\\

SubSIM
    & 7.92$_{\pm0.12}$ & 20.78$_{\pm0.39}$ & 37.09$_{\pm0.34}$ 
    & \underline{3.99}$_{\pm0.11}$ & 15.08$_{\pm0.15}$ & 28.71$_{\pm0.39}$ 
    & 4.28$_{\pm0.13}$ & 15.64$_{\pm0.20}$ & 28.41$_{\pm0.45}$ 
    & 17.45$_{\pm0.08}$ & 36.82$_{\pm0.14}$ & 78.70$_{\pm2.29}$\\

\midrule

PIANO
    & 2.43$_{\pm0.08}$ & 9.87$_{\pm0.32}$ & 17.36$_{\pm0.18}$ 
    & 2.76$_{\pm0.06}$ & 11.39$_{\pm0.23}$ & 20.58$_{\pm0.19}$ 
    & 3.85$_{\pm0.09}$ & 11.48$_{\pm0.18}$ & 22.48$_{\pm0.23}$ 
    & 16.93$_{\pm0.09}$ & 32.67$_{\pm0.11}$ & 41.82$_{\pm0.25}$\\
    
ToupleGDD
    & 2.39$_{\pm0.12}$ & 10.43$_{\pm0.38}$ & 18.65$_{\pm0.20}$ 
    & 2.83$_{\pm0.07}$ & 13.03$_{\pm0.10}$ & 25.05$_{\pm0.21}$ 
    & 3.89$_{\pm0.09}$ & 12.81$_{\pm0.15}$ & 24.19$_{\pm0.33}$ 
    & 17.02$_{\pm0.09}$ & 36.75$_{\pm0.25}$ & 61.69$_{\pm2.02}$\\

\midrule

IMINFECTOR
    & 1.77$_{\pm0.08}$ & 6.11$_{\pm0.12}$ & 12.58$_{\pm0.12}$
    & 1.69$_{\pm0.04}$ & 7.18$_{\pm0.11}$ & 15.87$_{\pm0.15}$ 
    & 1.32$_{\pm0.02}$ & 5.94$_{\pm0.06}$ & 12.49$_{\pm0.15}$ 
    & 1.21$_{\pm0.07}$ & 7.84$_{\pm0.03}$ & 15.79$_{\pm0.04}$\\

SGNN
    & 1.32$_{\pm0.05}$ & 7.14$_{\pm0.08}$ & 14.14$_{\pm0.23}$
    & 1.58$_{\pm0.05}$ & 7.66$_{\pm0.06}$ & 15.52$_{\pm0.14}$ 
    & 1.27$_{\pm0.04}$ & 6.48$_{\pm0.04}$ & 12.58$_{\pm0.06}$ 
    & 1.36$_{\pm0.04}$ & 8.16$_{\pm0.12}$ & 14.70$_{\pm0.03}$\\

GCNM
    & 1.10$_{\pm0.02}$ & 5.13$_{\pm0.03}$ & 10.90$_{\pm0.07}$
    & 1.29$_{\pm0.03}$ & 6.48$_{\pm0.08}$ & 12.43$_{\pm0.13}$ 
    & 1.12$_{\pm0.05}$ & 6.83$_{\pm0.73}$ & 12.10$_{\pm0.64}$ 
    & 1.27$_{\pm0.02}$ & 7.31$_{\pm0.06}$ & 13.50$_{\pm0.14}$\\
    
DeepIM
    & 7.66$_{\pm0.24}$ & 22.00$_{\pm0.41}$ & 38.80$_{\pm0.57}$
    & 3.78$_{\pm0.12}$ & 15.95$_{\pm0.27}$ & 28.98$_{\pm0.29}$ 
    & 2.63$_{\pm0.64}$ & 11.32$_{\pm0.56}$ & 25.49$_{\pm1.01}$ 
    & 16.92$_{\pm0.05}$ & 55.36$_{\pm0.81}$ & 76.11$_{\pm0.33}$\\

DeepSN
    & 1.42$_{\pm0.04}$ & 8.30$_{\pm0.10}$ & 14.69$_{\pm0.13}$
    & 1.84$_{\pm0.05}$ & 7.99$_{\pm0.12}$ & 16.22$_{\pm0.21}$ 
    & 1.55$_{\pm0.07}$ & 6.98$_{\pm0.22}$ & 15.95$_{\pm1.31}$ 
    & 1.45$_{\pm0.03}$ & 7.85$_{\pm0.02}$ & 16.72$_{\pm0.19}$\\
    
\midrule
HIM$_{MD}$
    & \underline{8.12}$_{\pm0.13}$ & \underline{23.82}$_{\pm0.61}$ & \underline{39.52}$_{\pm0.53}$ 
    & 3.62$_{\pm0.21}$ & \underline{16.26}$_{\pm0.20}$ & \underline{29.29}$_{\pm0.27}$ 
    & \underline{5.91}$_{\pm0.10}$ & \underline{18.37}$_{\pm0.28}$ & \underline{34.12}$_{\pm0.55}$ 
    & \underline{17.74}$_{\pm0.10}$ & \underline{57.02}$_{\pm0.59}$ & \underline{80.58}$_{\pm0.97}$\\  

HIM
    & \textbf{8.27}$_{\pm0.18}$ & \textbf{24.22}$_{\pm0.63}$ & \textbf{40.43}$_{\pm0.83}$
    & 3.89$_{\pm0.11}$ & \textbf{17.17}$_{\pm0.19}$ & \textbf{30.76}$_{\pm0.33}$ 
    & \textbf{6.14}$_{\pm0.10}$ & \textbf{18.79}$_{\pm0.30}$ & \textbf{34.51}$_{\pm0.63}$ 
    & \textbf{18.09}$_{\pm0.04}$ & \textbf{57.91}$_{\pm0.64}$ & \textbf{81.61}$_{\pm1.43}$\\ 

\bottomrule
\end{tabular}}
\label{table:result-lt}
\end{center}
\vspace{-2em}
\end{table*}
\subsubsection{Results with Known Diffusion Parameters}
Table~\ref{table:norm-ic} and Table~\ref{table:norm-lt} present the results for the standard IM problem under the IC and LT models. 

\textbf{Results under the IC}.
Traditional methods, including IMM and SubSIM, serve as reference baselines for evaluating performance under standard IM settings with known diffusion models.
IMM and SubSIM generally outperform learning-based baselines under the IC model. 
Both methods offer a $(1 - 1/e - \epsilon)$ approximation guarantee when the diffusion model and its parameters are completely known.
They rely on this knowledge and repeatedly simulate diffusion processes to obtain effective solutions.
PIANO and ToupleGDD also rely on known diffusion models to design reward functions for reinforcement learning, and they show competitive performance across four datasets.
In contrast, other learning-based methods might not provide such theoretical guarantees but demonstrate effective empirical performance.
Although our method performs slightly below the two traditional approaches, the results remain comparable and even surpass them in some cases. 
This demonstrates the effectiveness of our method as a novel framework for solving the standard IM problem.
More importantly, our methods achieve superior performance compared to other graph learning-based approaches. 
IMINFECTOR and SGNN use individual diffusion cascades as pseudo-labels, while GCNM generates them based on node degree information. 
Moreover, DeepIM and DeepSN learn from seed users and their final influence spread. The lack of explicit propagation process modeling and limited expressive power constrain their performance, indicating a failure to capture the complex patterns of influence spread.
The results of \HIMMD~demonstrate that our representation learning framework effectively captures social influence features from social data, particularly the influence propagation instances, using learned hyperbolic representations.
This also validates the feasibility of using hyperbolic geometry to estimate influence strength for IM tasks.
HIM further improves performance by incorporating our proposed ASW algorithm, showing that leveraging distances between user representations can better capture influence dynamics and lead to more effective solutions.

\textbf{Results under the LT}.
Under the LT model, the performance of different methods varies significantly. 
The two traditional methods, IMM and SubSIM, still perform strongly across all settings. 
However, most learning-based methods, except DeepIM, show limited effectiveness across four datasets. 
This indicates a lack of generalization ability and highlights their inability to capture meaningful influence patterns in the many-to-one propagation setting under the LT model.
In contrast, both of our methods, \HIMMD~and HIM, achieve strong results. 
The results of \HIMMD~demonstrate that our approach can effectively estimate users' influence strength to solve the IM problem under the LT model, highlighting the generalization capability of our learning strategy.
Moreover, HIM achieves superior results in numerous cases compared to the traditional methods, further validating the potential of leveraging hyperbolic geometry to infer social influence from data, especially the propagation history, in IM tasks.

Overall, the results under the IC and LT models confirm the effectiveness of the proposed method in standard IM problems with fully known diffusion models.

\subsubsection{Results with Unknown Diffusion Parameters}
Table~\ref{table:result-ic} and Table~\ref{table:result-lt} present the results under two diffusion model variants: WIC and WLT.
It is worth noting that IM problems under unknown diffusion model parameters are more challenging.

\textbf{Results under the WIC}.
Without knowing the propagation parameters, traditional methods cannot guarantee their approximation bounds.
IMM and SubSIM fail to sample reverse reachable sets that accurately reflect influence spread, leading to a significant drop in performance.
PIANO and ToupleGDD also exhibit limited performance when the diffusion model parameters are unknown.
Performance varies significantly among graph learning-based baselines. 
Similar to the results under the IC model, the inherent limitations of these methods also restrict their performance when diffusion parameters are unknown.
Instead, our methods consistently outperform all baselines when the diffusion parameters are unknown.
Compared to graph learning-based methods, our methods achieve significant performance improvements.
The performance gain primarily stems from our method’s ability to leverage hyperbolic representations to effectively extract social influence patterns from historical propagation processes, thereby enabling a more accurate identification of highly influential seed users.
Compared to HIM$_{MD}$, HIM achieves better performance, demonstrating the effectiveness of the proposed adaptive seed selection in exploring more potential candidates.
The experimental results demonstrate that our methods can effectively identify highly influential users under the complex WIC model.

\textbf{Results under the WLT}.
Similar to the LT model, the WLT model represents a more complex scenario with a many-to-one propagation pattern. 
Compared to LT, selecting target seed users under WLT is more challenging due to the uncertainty of user influence weights.
IMM and SubSIM have limited performance since they cannot perceive the different influence weights of users when sampling reverse reachable sets.
For the same reason, PIANO and ToupleGDD fail to design effective rewards, leading to even worse performance.
Similar to the results under the LT model, most graph learning-based methods, except DeepIM, struggle with the propagation mode and fail to function effectively under the WLT model.
DeepIM demonstrates competitive performance in certain cases.
In comparison, our methods consistently outperform all compared methods in nearly all cases.
The results demonstrate the robustness of our method in learning propagation patterns for seed user selection under stochastic influence weights.
Moreover, the strong performance of our method further confirms its effectiveness in capturing social influence patterns during propagation via hyperbolic representations, demonstrating excellent generalization to complex diffusion scenarios. 

Overall, the results obtained under the WIC and WLT models demonstrate that our method does not rely on any assumptions about diffusion parameters, highlighting its strong potential for real-world influence propagation scenarios.

\subsection{Efficiency Analysis}
\begin{figure}[!ht]
\vspace{-1em}
\centering
    \subfloat{\includegraphics[width=0.38\columnwidth]{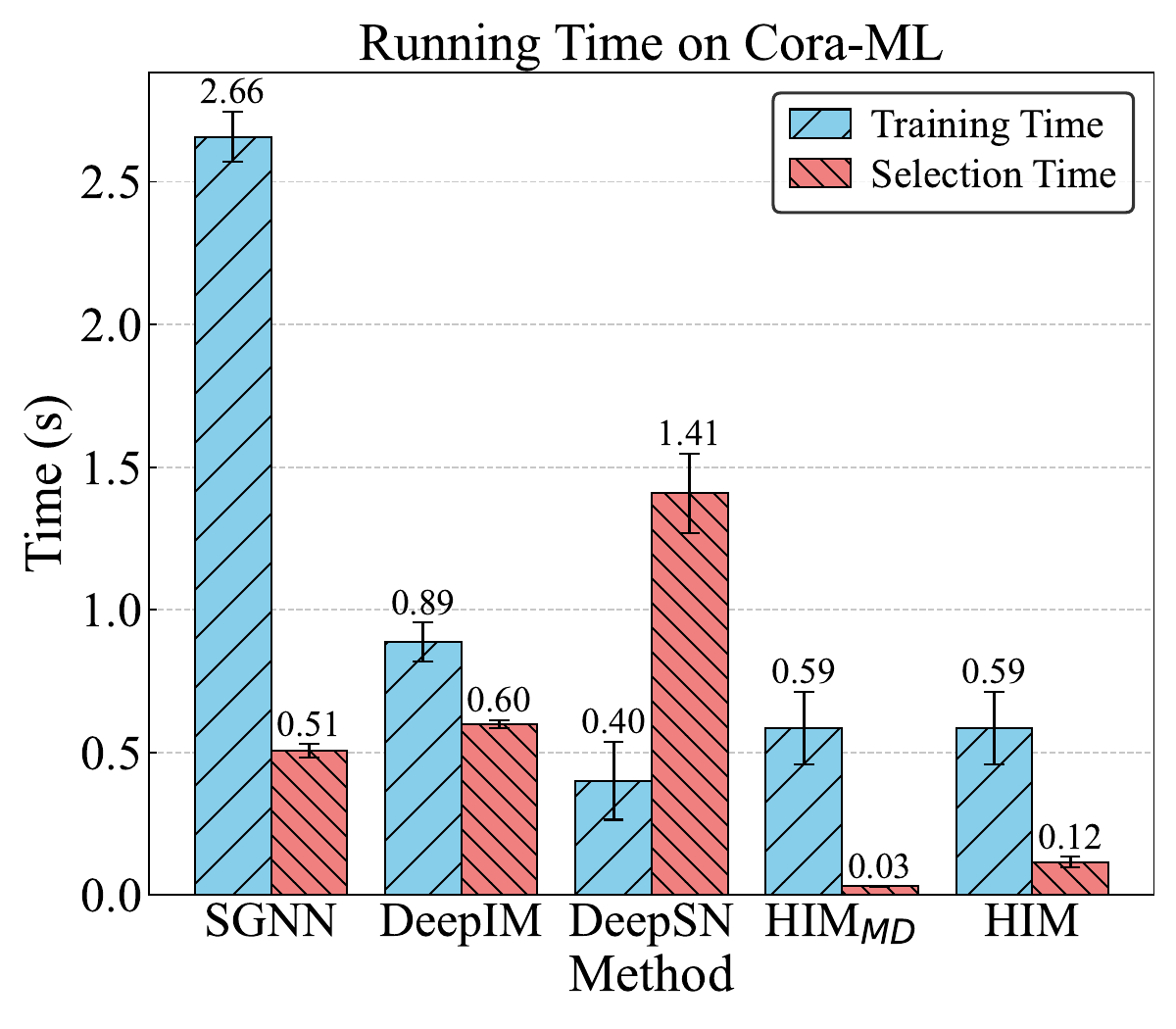}}
    \subfloat{\includegraphics[width=0.38\columnwidth]{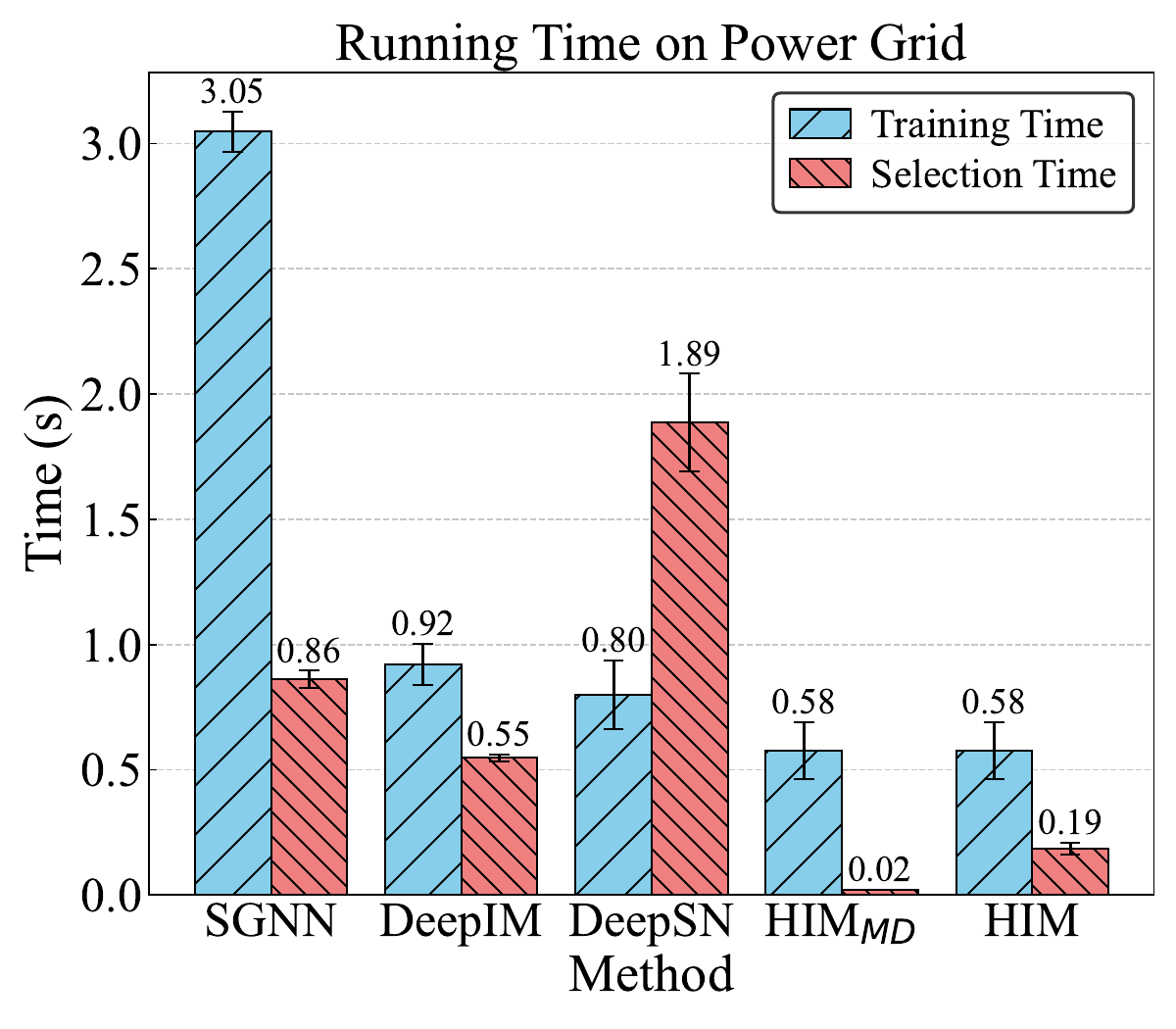}}

    \subfloat{\includegraphics[width=0.38\columnwidth]{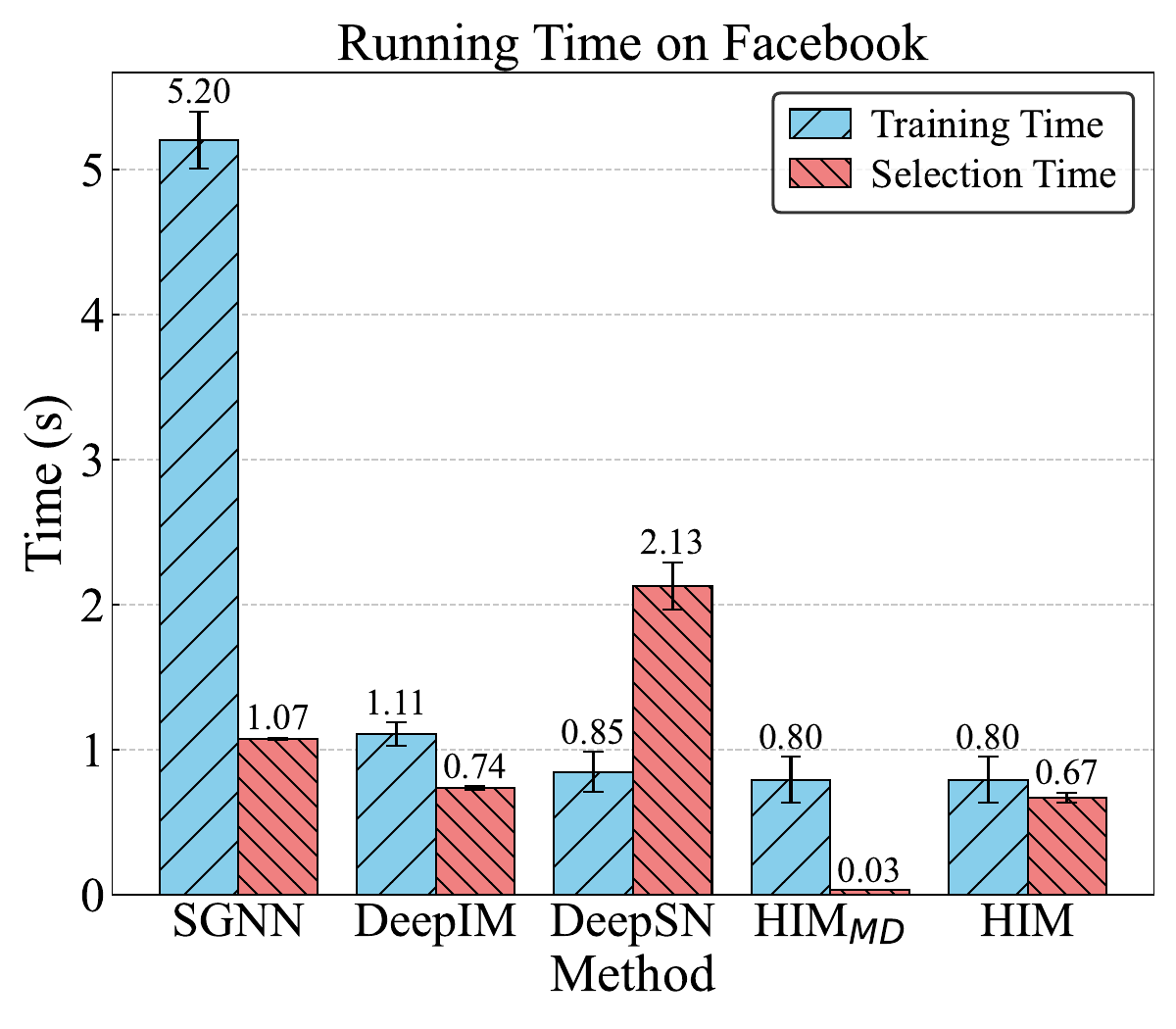}}
    \subfloat{\includegraphics[width=0.38\columnwidth]{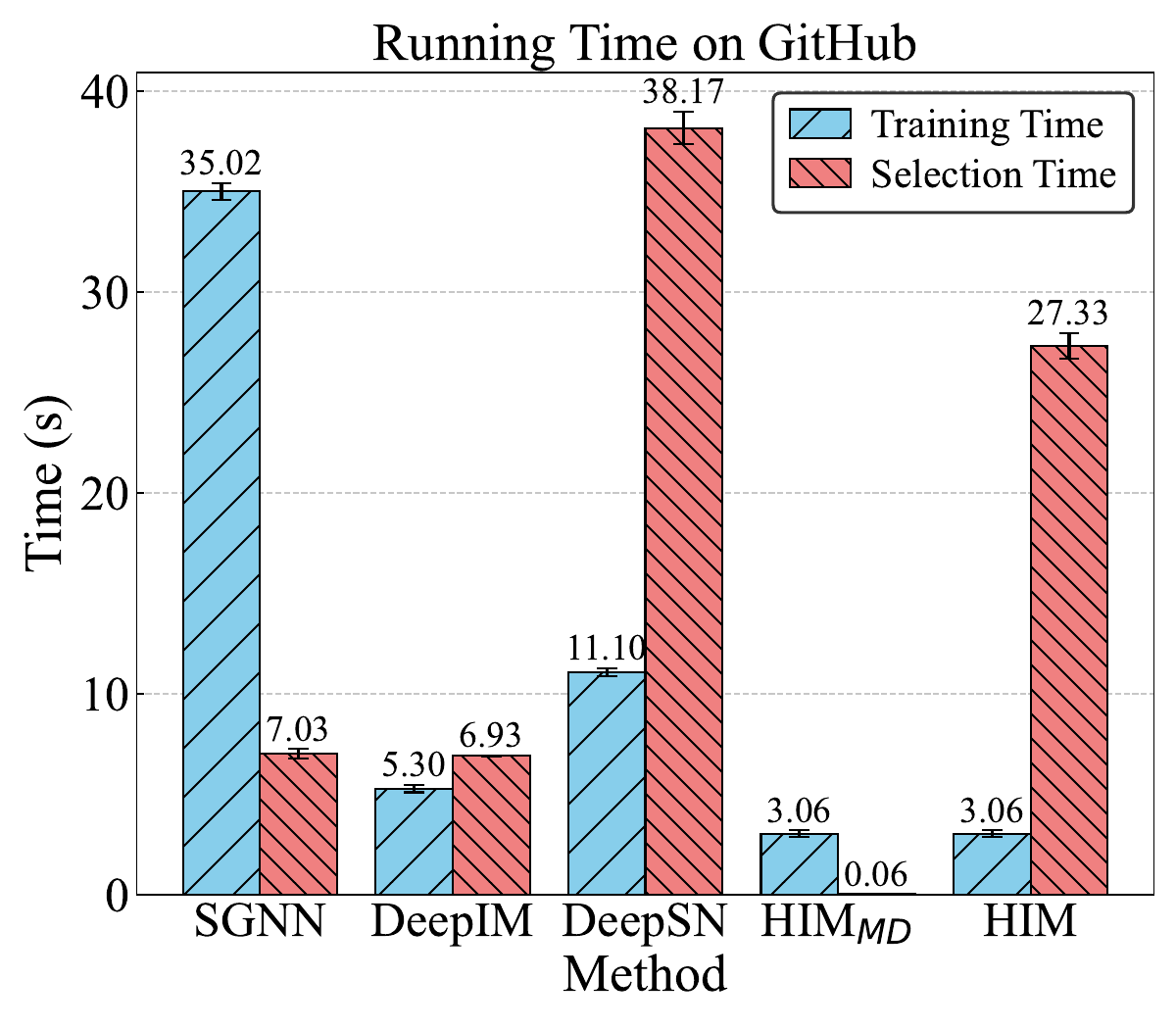}}
    \caption{Results of training and selection time comparison.}
    \label{fig:time}
\end{figure}
We compare the time efficiency of our method with other baselines that use representation learning as the core component across four datasets.
We report the results under the WIC model since the running times are almost consistent across different diffusion models.
We perform 10 runs of the training and selection process to report the average training time per epoch and the average selection time, both measured in seconds.
To better present the results, we omit IMINFECTOR and GCNM because they run much slower than the other methods.
The results are shown in Figure~\ref{fig:time}.

Overall, our methods demonstrate superior training and selection efficiency.
Except for the smallest dataset, Cora-ML, where our training time is slightly slower than DeepSN, our method consistently shows better training efficiency on all other datasets. 
As the network size increases, the training efficiency advantage of our method becomes increasingly evident.
For the selection time, \HIMMD shows a significant advantage, as its computational cost mainly stems from a sorting operation. 
Although HIM incurs additional overhead on the GitHub dataset, it exhibits overall efficiency across all datasets. 
Later, we demonstrate that these baselines struggle to handle large-scale networks, whereas HIM maintains both high effectiveness and efficiency as the network size increases.

\subsection{Performance on Large-scale Network}
\begin{figure}[!ht]
\vspace{-2em}
\centering
    \subfloat{\includegraphics[width=0.40\columnwidth]{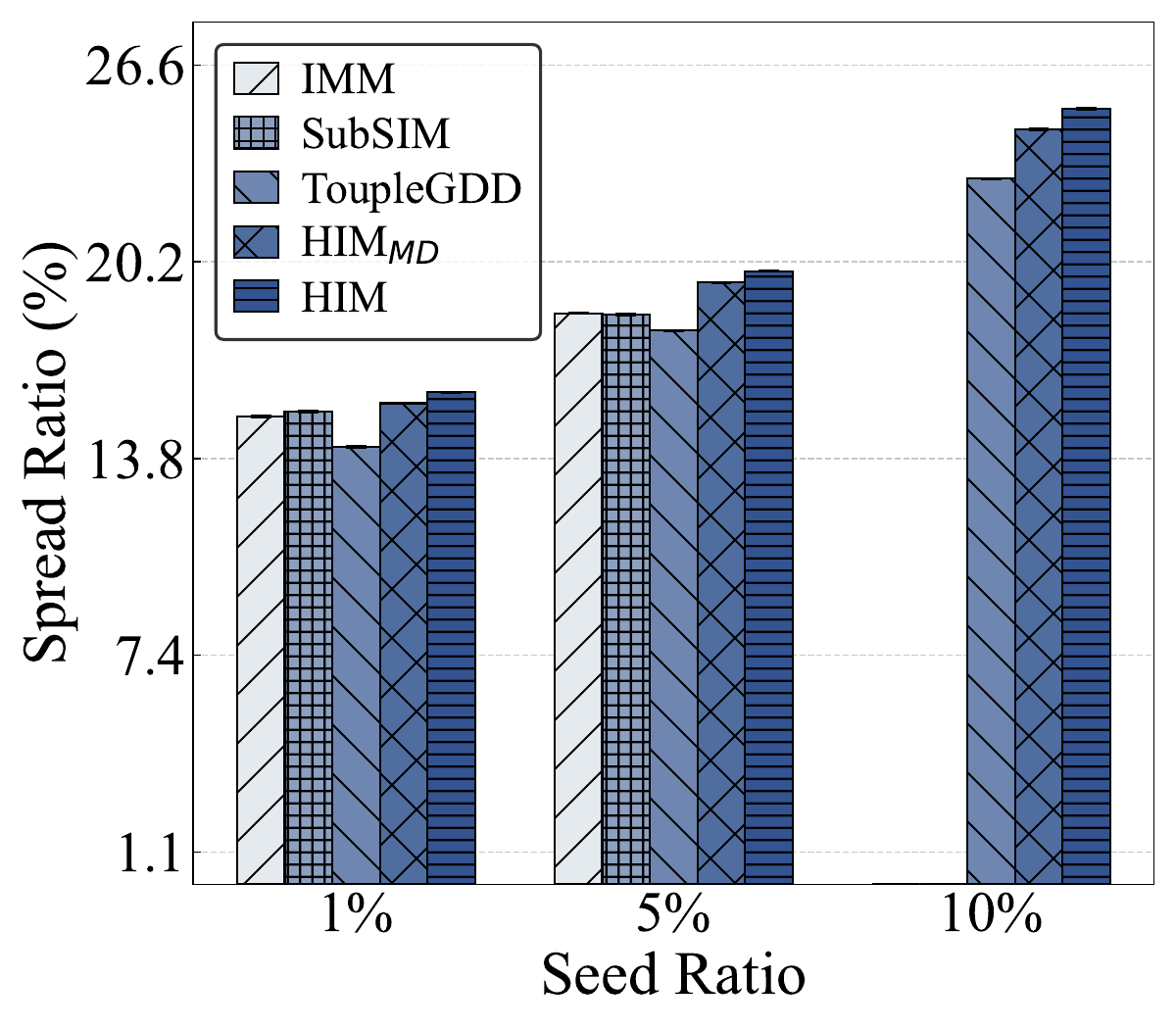}}
    \subfloat{\includegraphics[width=0.40\columnwidth]{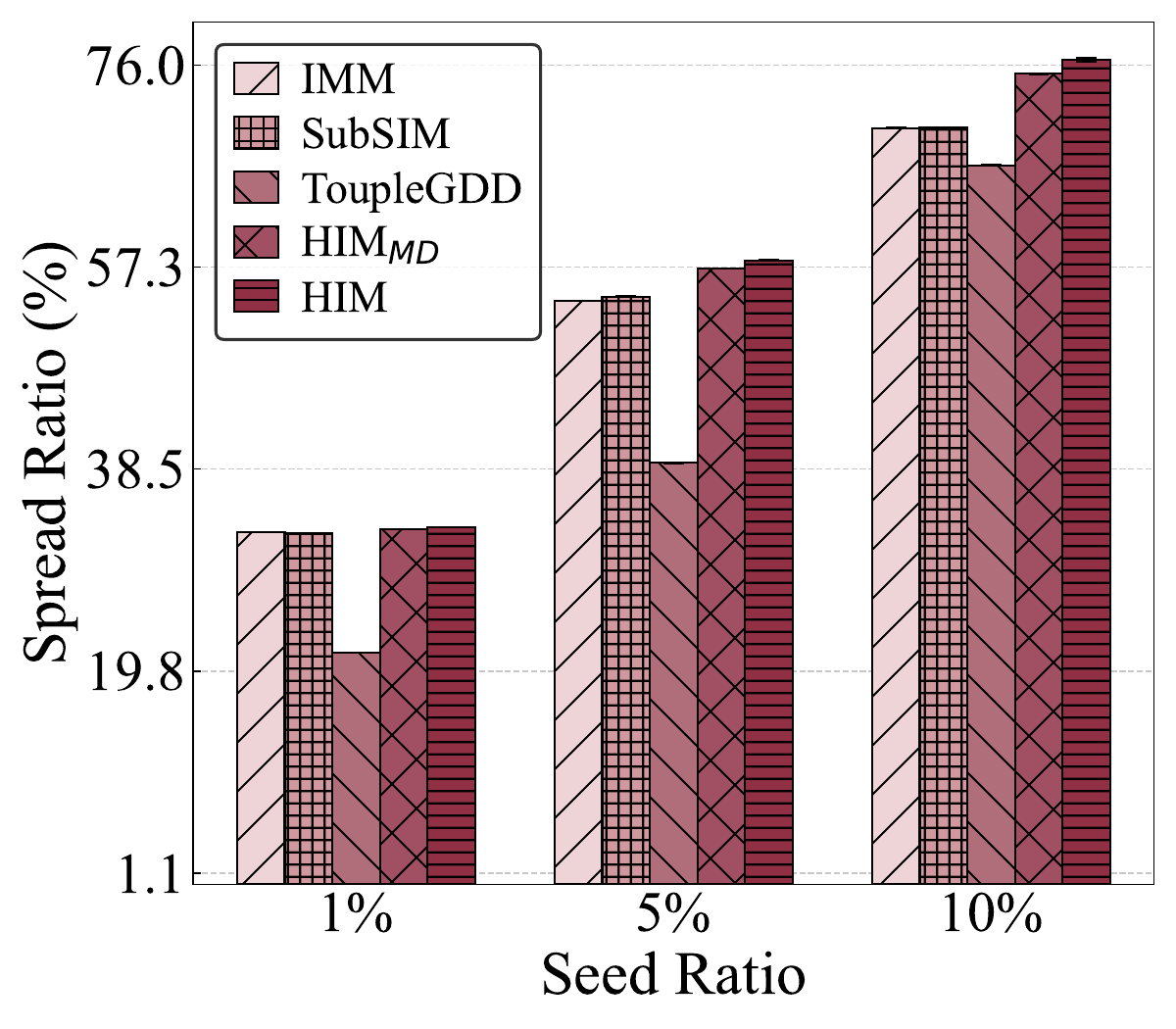}}
    \caption{Results on Youtube under WIC (left) and WLT (right).}
    \label{fig:large}
\end{figure}
We investigate the scalability of \HIMMD~and HIM on a large-scale network, YouTube, with millions of nodes.
The results under the WIC and WLT models are shown in Figure~\ref{fig:large}.
We compare our method with IMM, SubSIM, and ToupleGDD. 
Other learning-based baselines, including PIANO, IMINFENCTOR, SGNN, GCNM, DeepIM, and DeepSN, could not be executed in our experimental environment due to memory constraints, indicating practical scalability limitations in large-scale settings.
The results show that our method consistently outperforms IMM, SubSIM, and ToupleGDD, demonstrating superior effectiveness and scalability.
IMM and SubSIM fail to produce valid results within 24 hours at a 10\% seed ratio under the WIC model. 
As $k$ increases, the unknown diffusion parameters affect the RR set sampling, resulting in high computational cost.

In terms of running efficiency, IMM and SubSIM require over 8 and 18 hours, respectively, to produce results at 1\% and 5\% seed ratios under the WIC model.
ToupleGDD requires a minimum selection time of more than 2 hours at a 1\% seed ratio.
Instead, the maximum selection time for \HIMMD~and HIM is 2 seconds and 20 minutes, respectively, across all cases.
Besides, it is worth noting that the total training time for our method is around 20 minutes, significantly shorter than that of ToupleGDD, which exceeds 15 hours.
Similar results are observed under both the IC and LT models.
The results demonstrate that HIM$_{MD}$ and HIM can effectively and efficiently identify high-influence seed users in a large-scale network with millions of nodes, highlighting their potential for practical large-scale IM scenarios.

\subsection{Ablation Study}
We validate the effectiveness of our method design through a comprehensive ablation study.
The ablation results on four datasets under the WIC and WLT models with seed ratio $10\%$ are shown in Table~\ref{table:ablation}.
(1) \textbf{w/o Hyp} represents the implementation of the entire HIM model in Euclidean space. 
The significant decrease in influence ratio across all cases indicates that, compared to Euclidean space, hyperbolic space more effectively captures influence spread patterns.
This also validates the effectiveness of hyperbolic network embedding for IM, leveraging the hierarchical structure of hyperbolic space to distinguish highly influential users.
(2) \textbf{w/o Rot} removes rotation operations from Eq. (\ref{eq:relation-score}) and Eq. (\ref{eq:propagation_score}).
Instead, we directly calculate the squared Lorentzian distance between $\mathbf{x}_u$ and $\mathbf{x}_v$.
The experimental results show that the rotation operation enhances representation learning by better preserving hierarchical social influence patterns.
As a result, the learned representations enable improved characterization of social influence strength.
(3) \textbf{w/o NSL} removes the network structure learning module, and (4) \textbf{w/o IPL} removes the influence propagation learning module.
The results (3) and (4) show that both network structure information and historical spread patterns in propagation instances are crucial for accurately quantifying influence spread. In comparison, propagation instances play a more significant role in measuring potential influence trends.
\begin{table}[!tb]
\caption{The results of ablation study in average spread ratio (\%) on four datasets with seed ratio 10\%.}
\resizebox{\columnwidth}{!} {
\begin{tabular}{l|lcccc}
\toprule
\multicolumn{1}{c}{~} & \multicolumn{1}{c}{Setting}         
  & Cora-ML                                    
  & Power Grid           
  & Facebook           
  & GitHub     
\\ 
\midrule
\multirow{5}{*}{WIC}   
& (1) w/o Hyp & 37.15$_{\pm0.39}$         
& 18.28$_{\pm0.44}$              
& 30.78$_{\pm0.33}$                 
& 32.33$_{\pm1.09}$              
\\
       
& (2) w/o Rot & 38.52$_{\pm0.67}$       
& 18.52$_{\pm0.32}$                  
& 32.12$_{\pm0.52}$                
& 35.17$_{\pm0.34}$              
\\
        
& (3) w/o NSL & 38.02$_{\pm0.58}$       
& 18.19$_{\pm0.43}$                
& 31.51$_{\pm0.34}$               
& 35.23$_{\pm0.74}$             
\\
       
& (4) w/o IPL & 35.20$_{\pm0.83}$         
& 17.43$_{\pm0.57}$                     
& 27.89$_{\pm0.58}$                   
& 28.59$_{\pm0.44}$                    
\\ \cline{2-6} 
           
& \multicolumn{1}{c}{HIM}      
& \textbf{38.87}$_{\pm0.96}$ 
& \textbf{19.13}$_{\pm0.50}$ 
& \textbf{32.69}$_{\pm0.59}$ 
& \textbf{36.27}$_{\pm0.22}$ 
\\ 
\midrule
\multirow{5}{*}{WLT}                    
& (1) w/o Hyp & 38.15$_{\pm0.75}$           
& 28.91$_{\pm0.42}$                       
& 32.02$_{\pm0.43}$                     
& 76.35$_{\pm0.55}$                
\\
& (2) w/o Rot & 40.19$_{\pm0.91}$        
& 30.51$_{\pm0.47}$                      
& 33.85$_{\pm0.80}$                      
& 80.34$_{\pm0.69}$                   
\\
& (3) w/o NSL & 39.27$_{\pm0.61}$  
& 30.35$_{\pm0.28}$                 
& 33.02$_{\pm0.50}$                    
& 79.82$_{\pm1.54}$                    
\\
& (4) w/o IPL & 37.23$_{\pm0.56}$   
& 27.77$_{\pm0.24}$                  
& 31.88$_{\pm0.47}$                 
& 73.48$_{\pm2.17}$               
\\ \cline{2-6} 
& \multicolumn{1}{c}{HIM}      
& \textbf{40.43}$_{\pm0.83}$
& \textbf{30.76}$_{\pm0.33}$ 
& \textbf{34.51}$_{\pm0.63}$ 
& \textbf{81.61}$_{\pm1.43}$ 
\\ 
\bottomrule
\end{tabular} }
\label{table:ablation} 
\vspace{-1em}
\end{table}

\vspace{-1em}
\begin{figure}[!ht]
\centering
    \subfloat{\includegraphics[width=0.36\columnwidth]{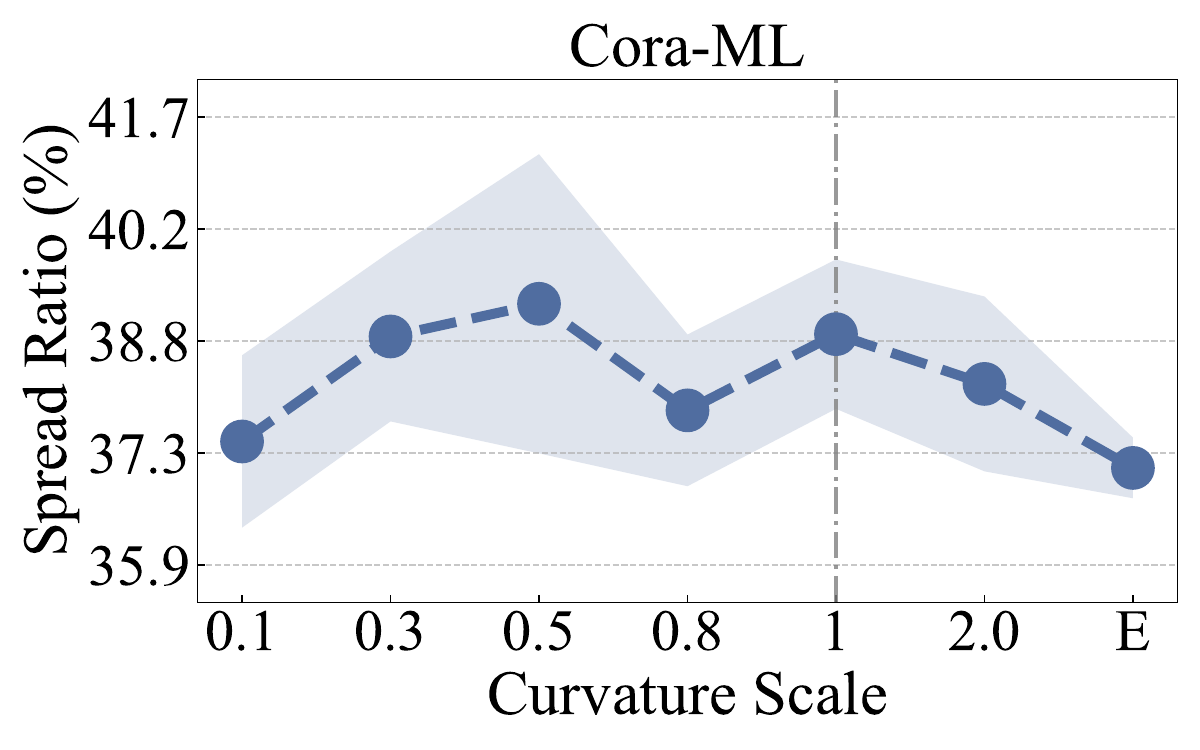}}
    \subfloat{\includegraphics[width=0.36\columnwidth]{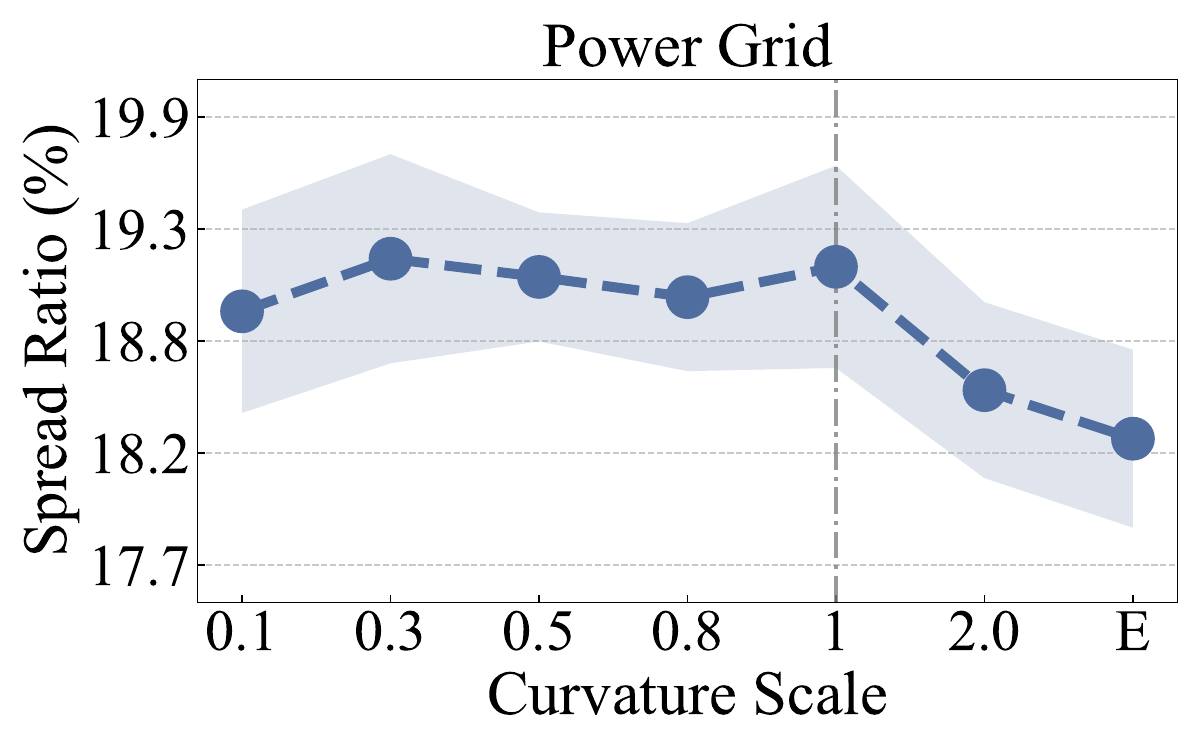}}
    \vspace{-1em}
    \subfloat{\includegraphics[width=0.36\columnwidth]{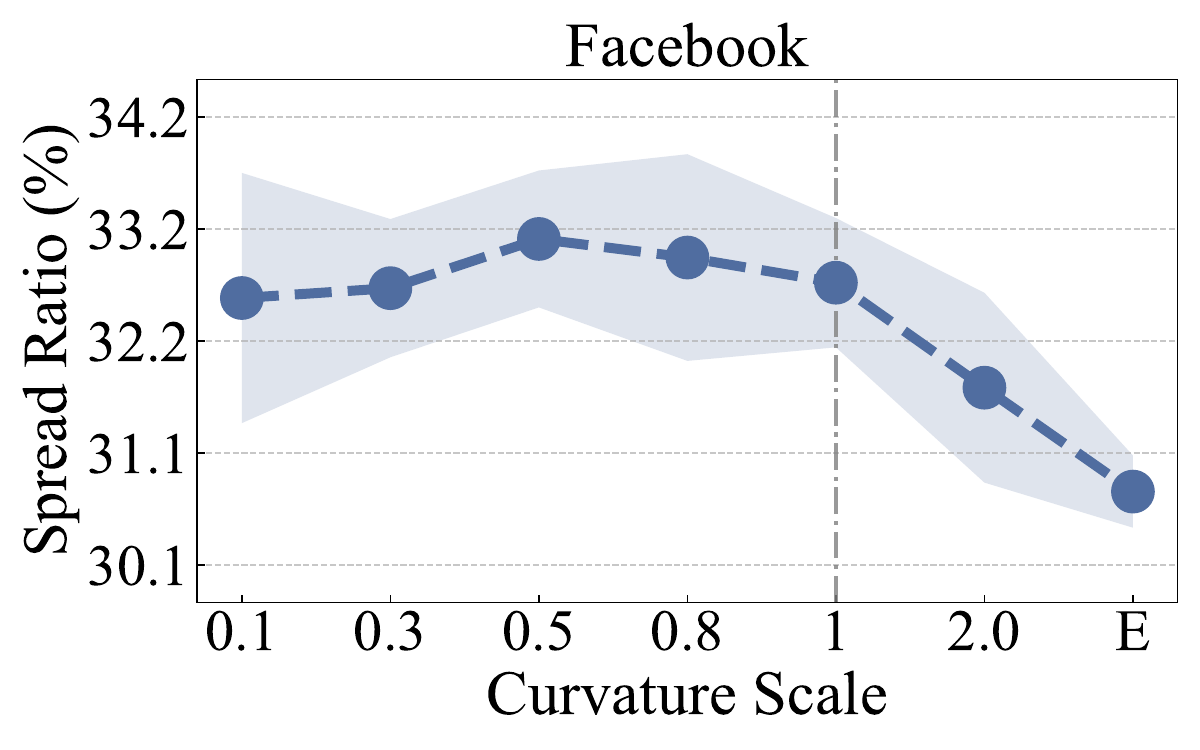}}
    \subfloat{\includegraphics[width=0.36\columnwidth]{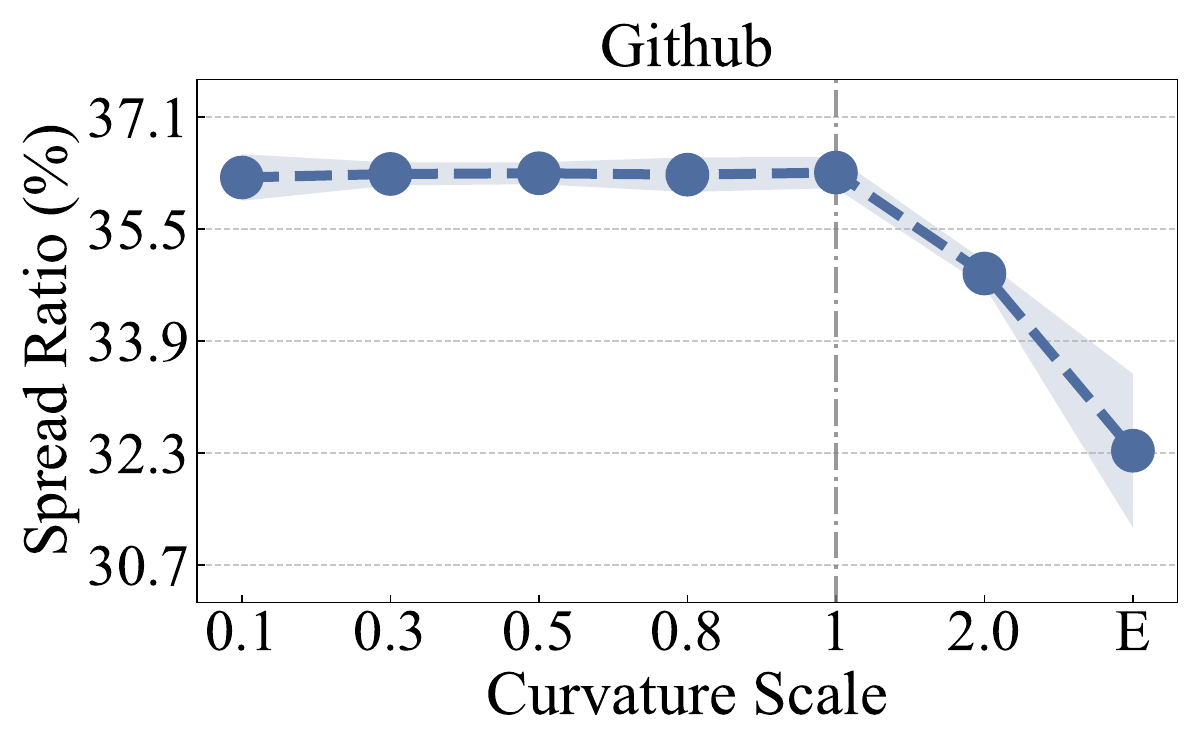}}
    \caption{Results on curvature scale under the WIC model.}
    \label{fig:c_wic}
    
    \subfloat{\includegraphics[width=0.36\columnwidth]{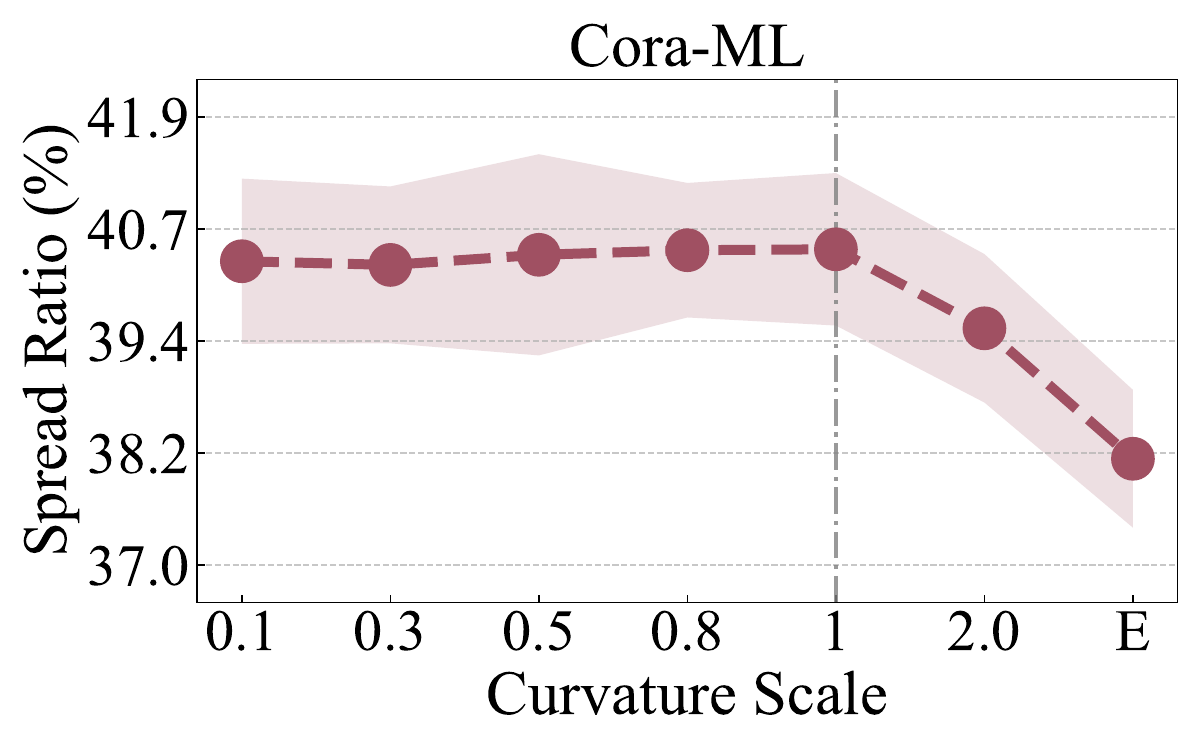}}
    \subfloat{\includegraphics[width=0.36\columnwidth]{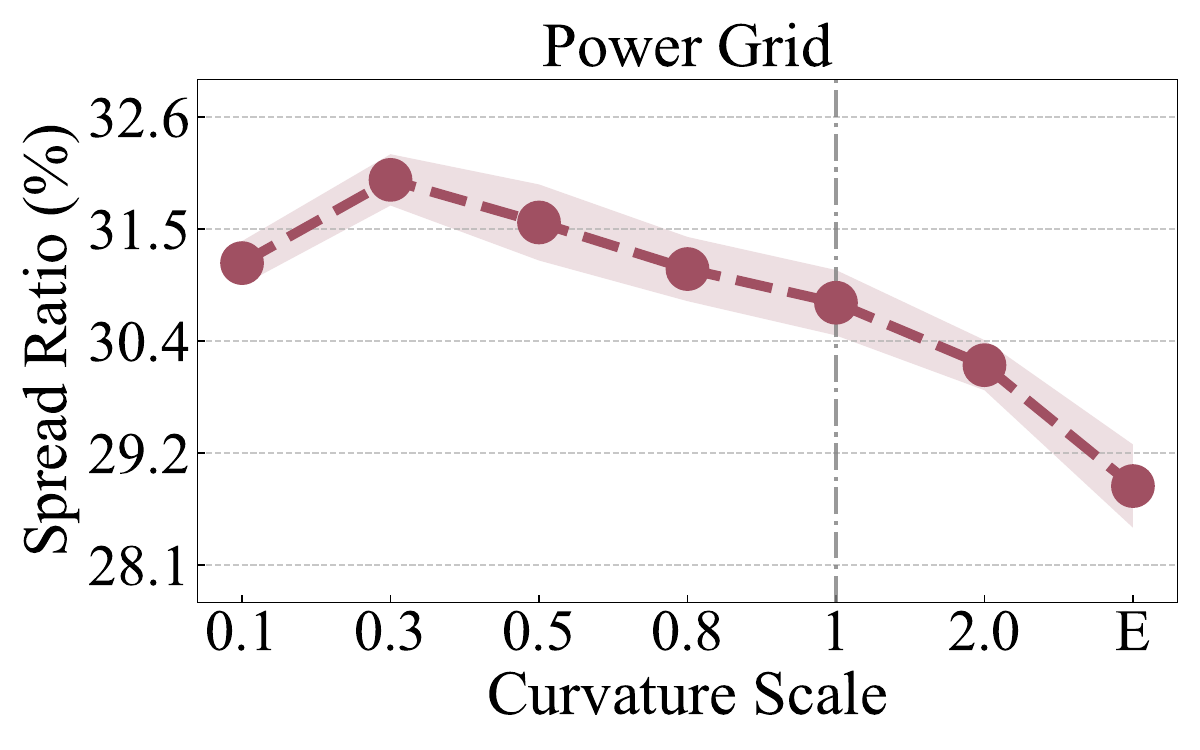}}
    \vspace{-1em}
    \subfloat{\includegraphics[width=0.36\columnwidth]{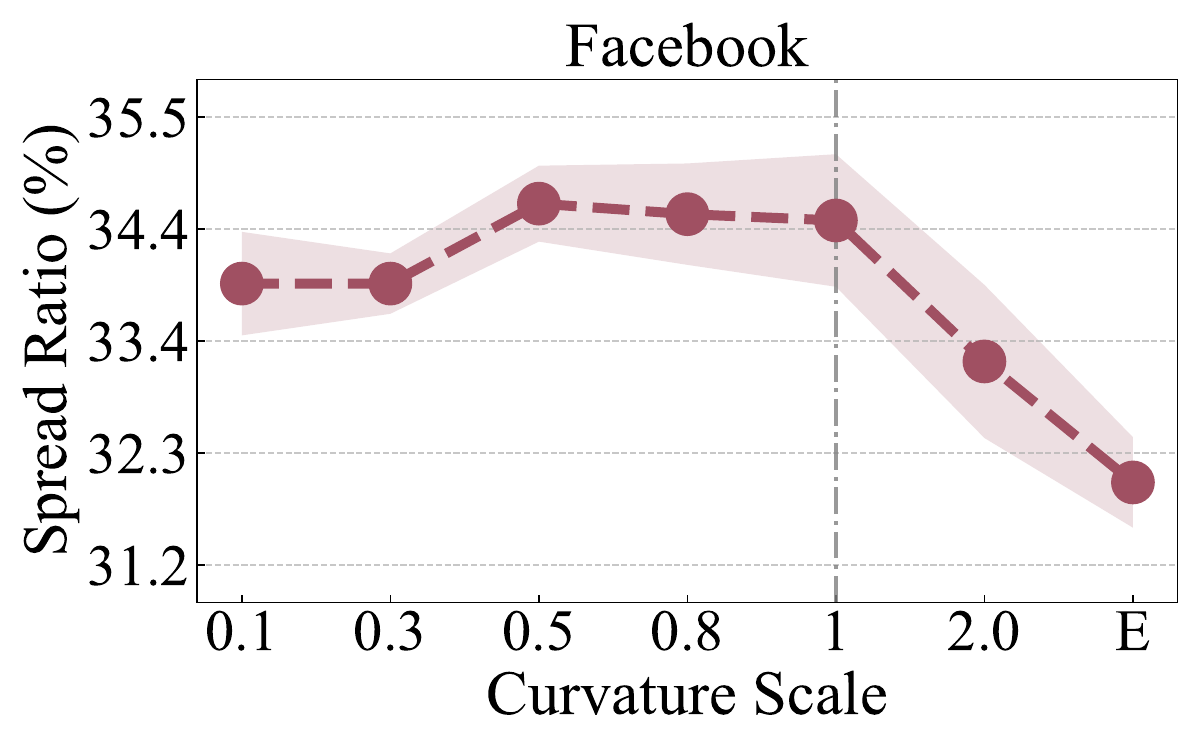}}
    \subfloat{\includegraphics[width=0.36\columnwidth]{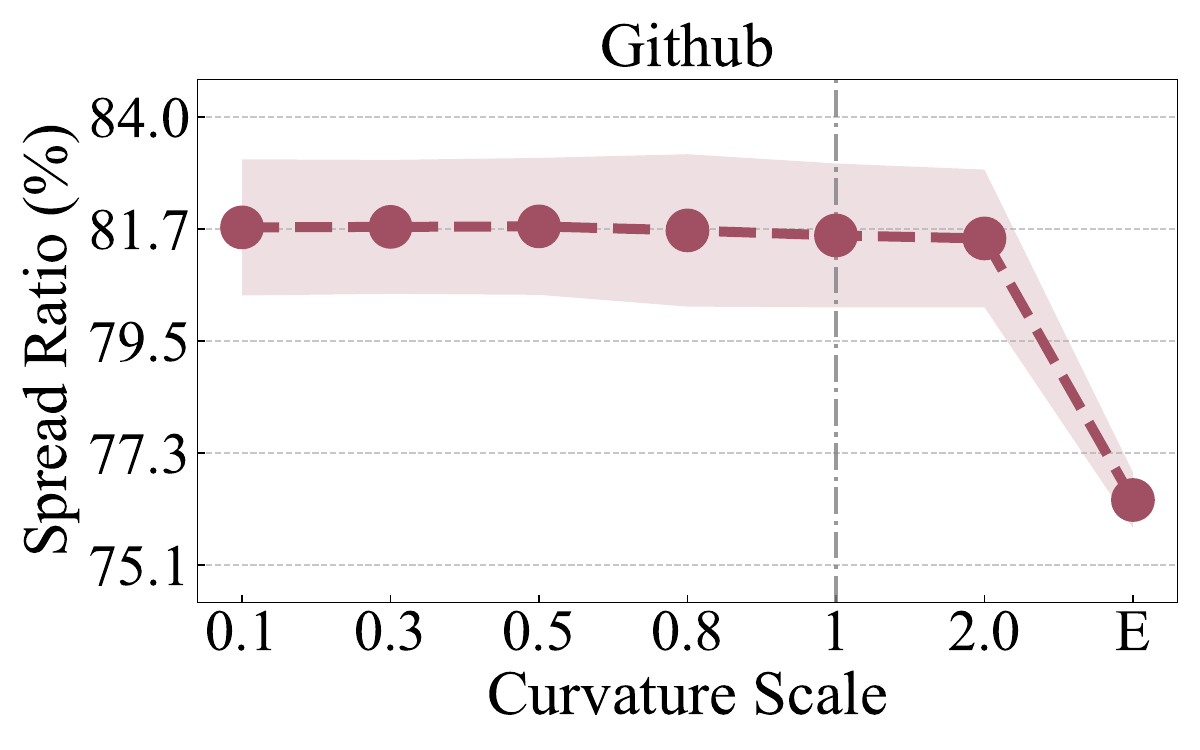}}
    \caption{Results on curvature scale under the WLT model.}
    \vspace{-2em}
    \label{fig:c_wlt}
\end{figure} 
\subsection{Parameter Sensitivity}
In this section, we analyze the parameter sensitivity of our method.
Specifically, we explore the effects of different curvature scales and the window coefficient on performance.

\textbf{Impact of Curvature}.
To study the impact of curvature, we investigate the effects of different curvature scales $\gamma$. 
The experimental results on four datasets under the WIC and WLT models are shown in Figures~\ref{fig:c_wic} and~\ref{fig:c_wlt}, respectively.
The seed ratio is $10\%$.
The vertical dashed line indicates the default $\gamma$.
Since the curvature is defined as $-1/\gamma$, a smaller $\gamma$ implies stronger negative curvature, while a larger $\gamma$ corresponds to weaker curvature. The Euclidean case (E) is included for reference.
Several observations can be made.
First, across all cases with $\gamma > 0$, the performance is consistently better than that in Euclidean space, suggesting that hyperbolic representations provide a more suitable geometric structure for capturing hierarchical social influence patterns in IM.
Second, the method's performance is not highly sensitive to the choice of curvature.
Within a reasonable range, performance nevertheless exhibits a mild upward trend as the magnitude of negative curvature increases (i.e., as $\gamma$ decreases, with $\gamma \le 1.0$).
Yet, when $\gamma$ becomes especially small (e.g., $\gamma = 0.1$), performance may start to decline. 
This suggests that excessively strong negative curvature might not yield further performance gains.
Overall, $\gamma = 1$ provides a reasonable and practical default setting.

\begin{figure}[!ht]
\vspace{-1em}
\centering
    \subfloat{\includegraphics[width=0.38\columnwidth]{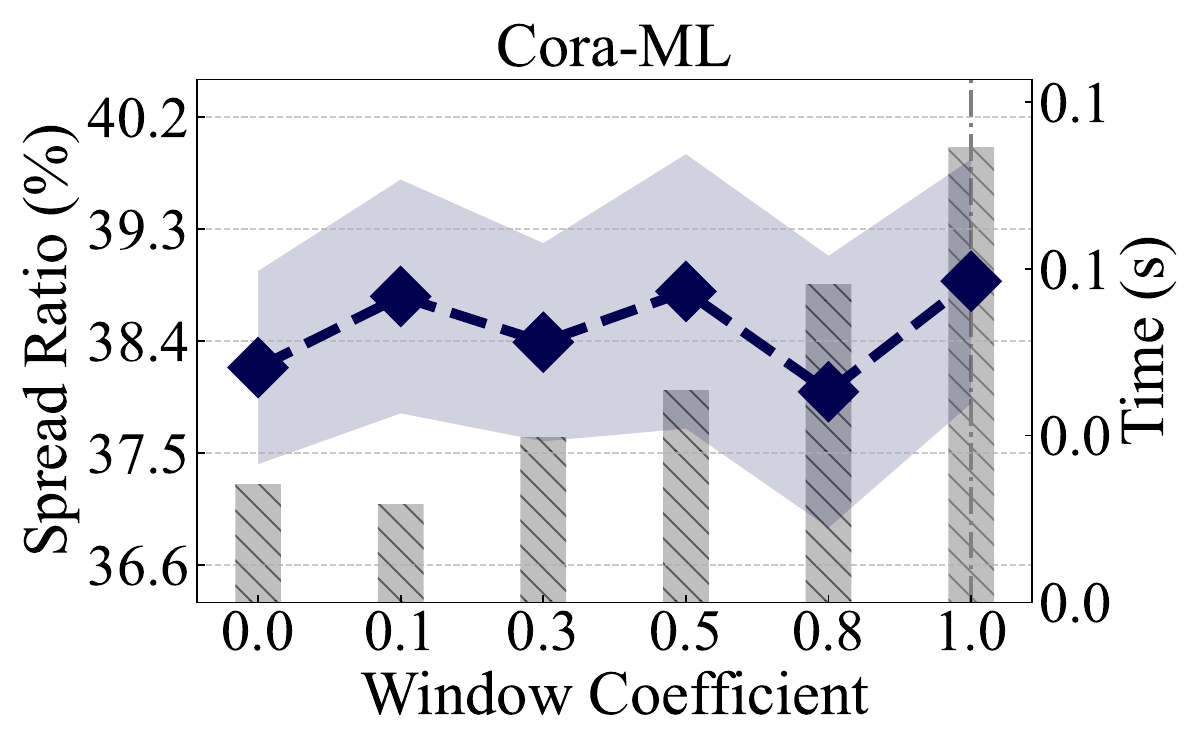}}
    \subfloat{\includegraphics[width=0.38\columnwidth]{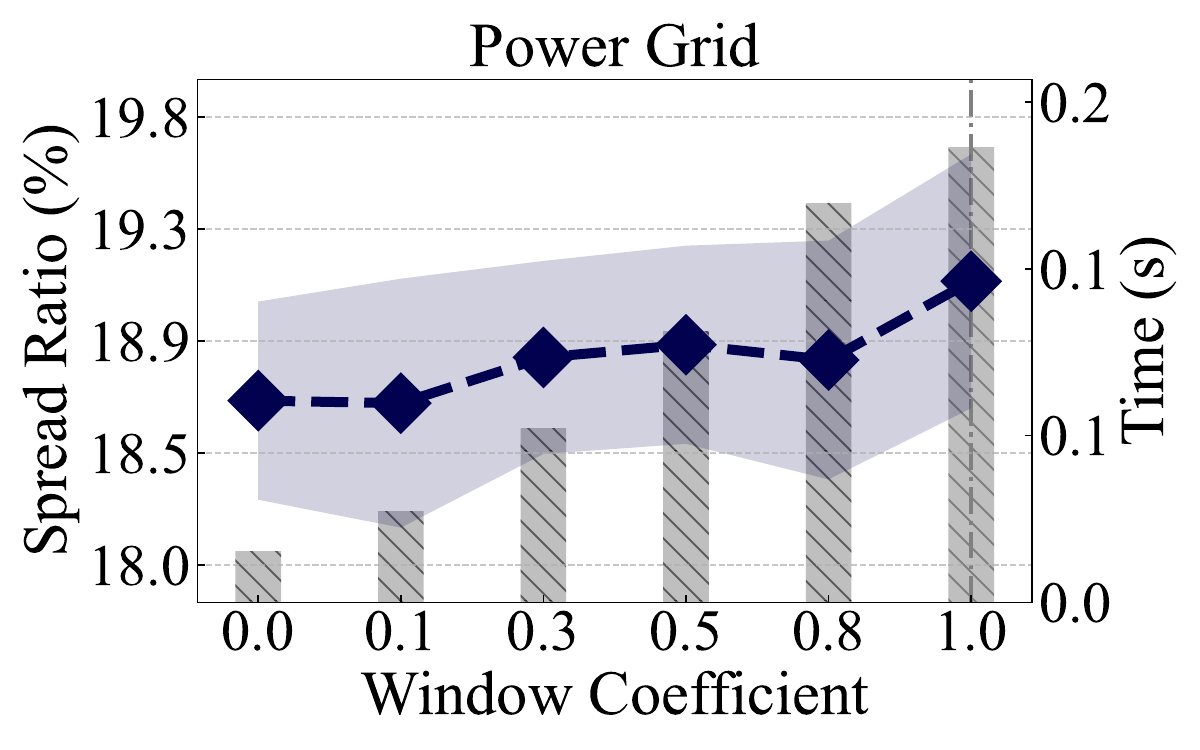}}
    \vspace{-1em}
    \subfloat{\includegraphics[width=0.38\columnwidth]{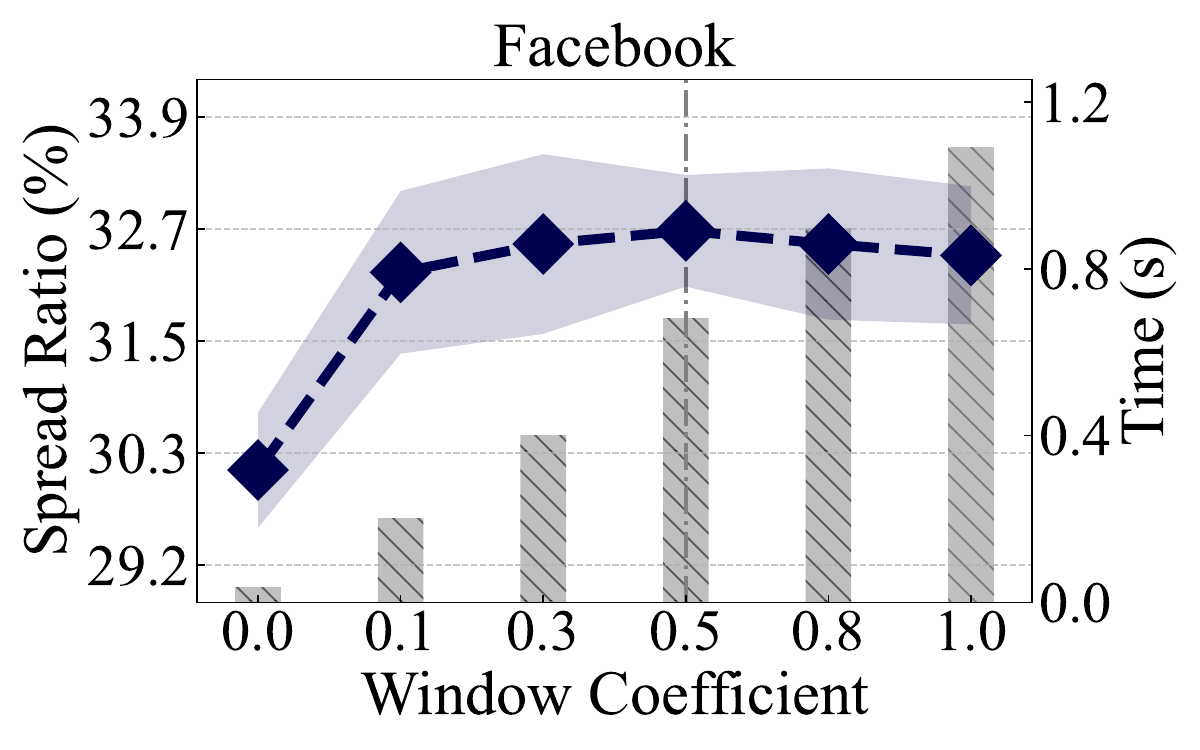}}
    \subfloat{\includegraphics[width=0.38\columnwidth]{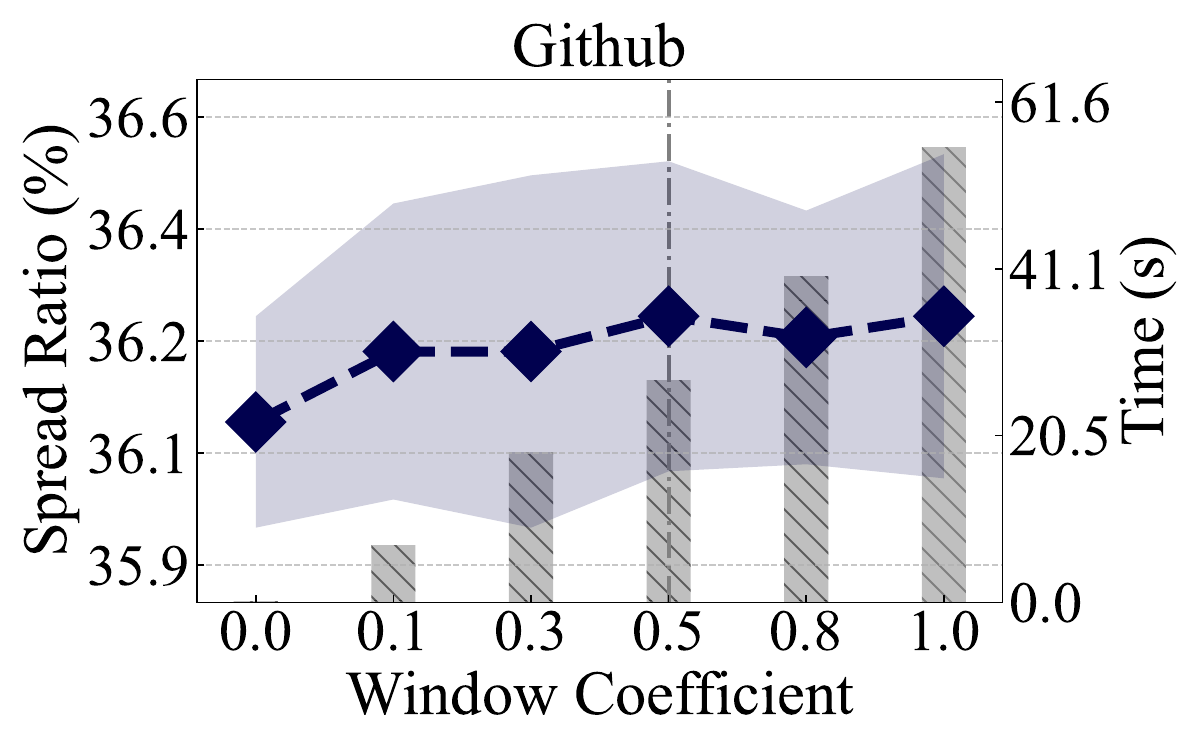}}
    \caption{Results on window coefficient under the WIC model.}
    \label{fig:w_wic}

    \subfloat{\includegraphics[width=0.38\columnwidth]{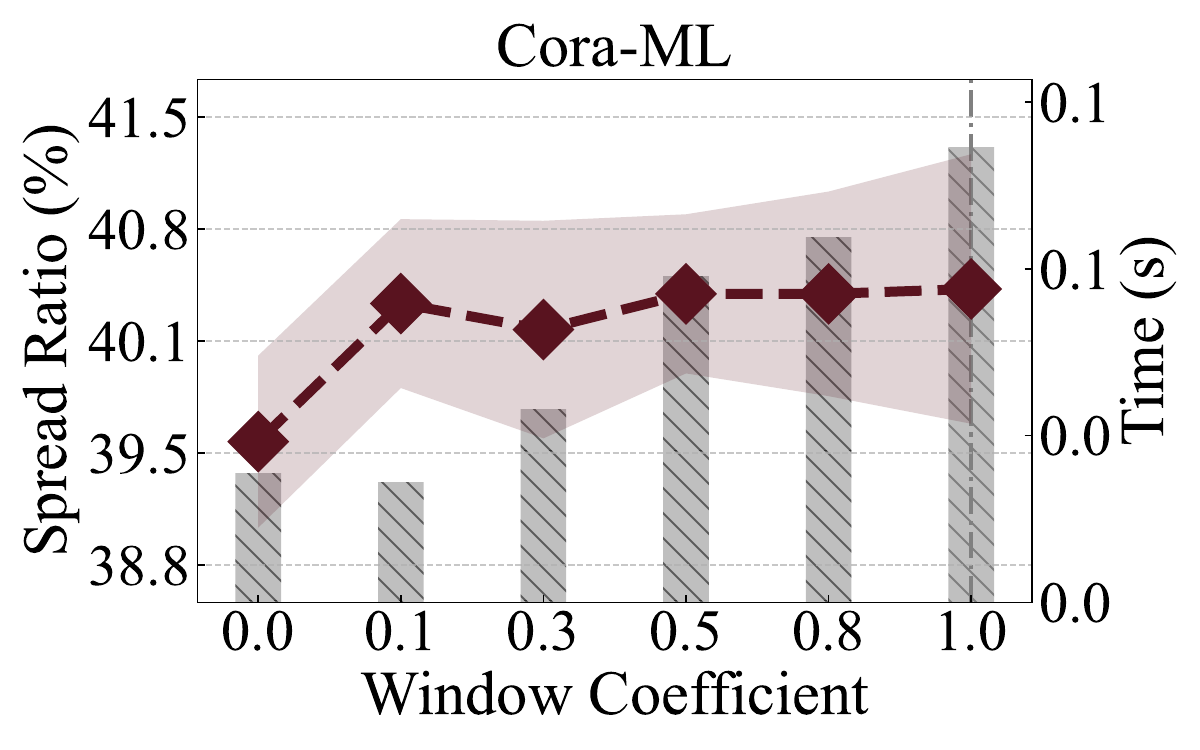}}
    \subfloat{\includegraphics[width=0.38\columnwidth]{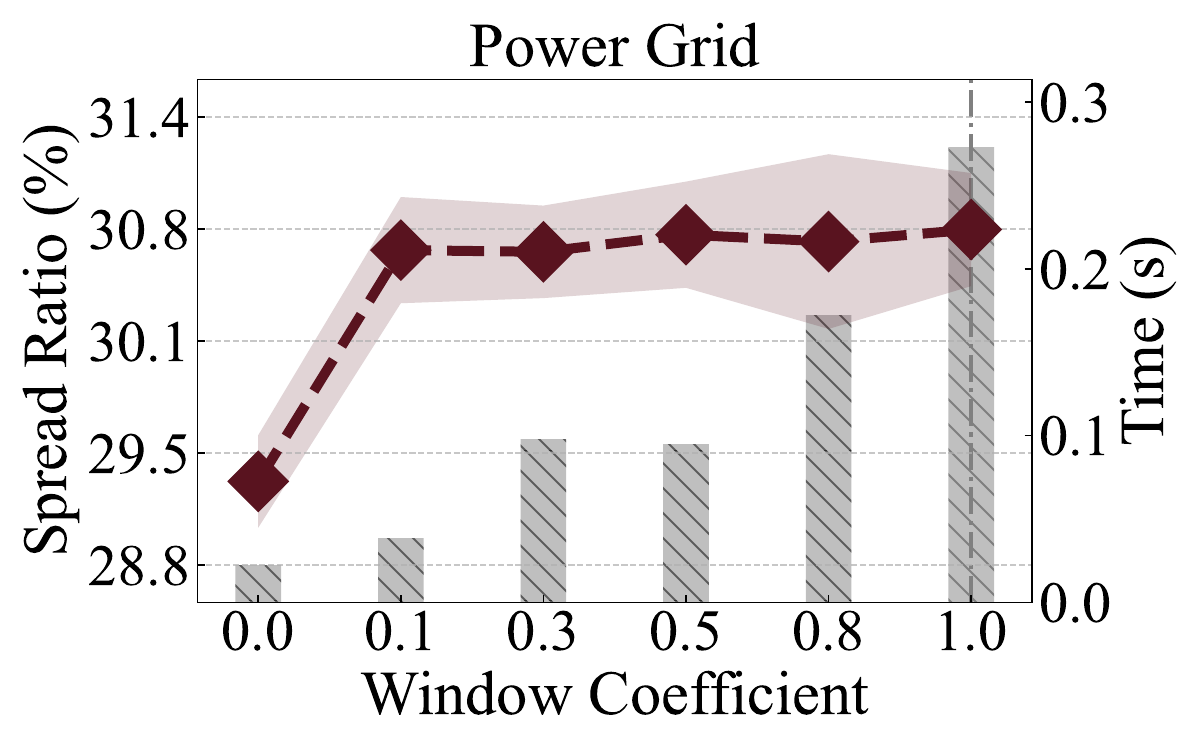}}
    \vspace{-1em}
    \subfloat{\includegraphics[width=0.38\columnwidth]{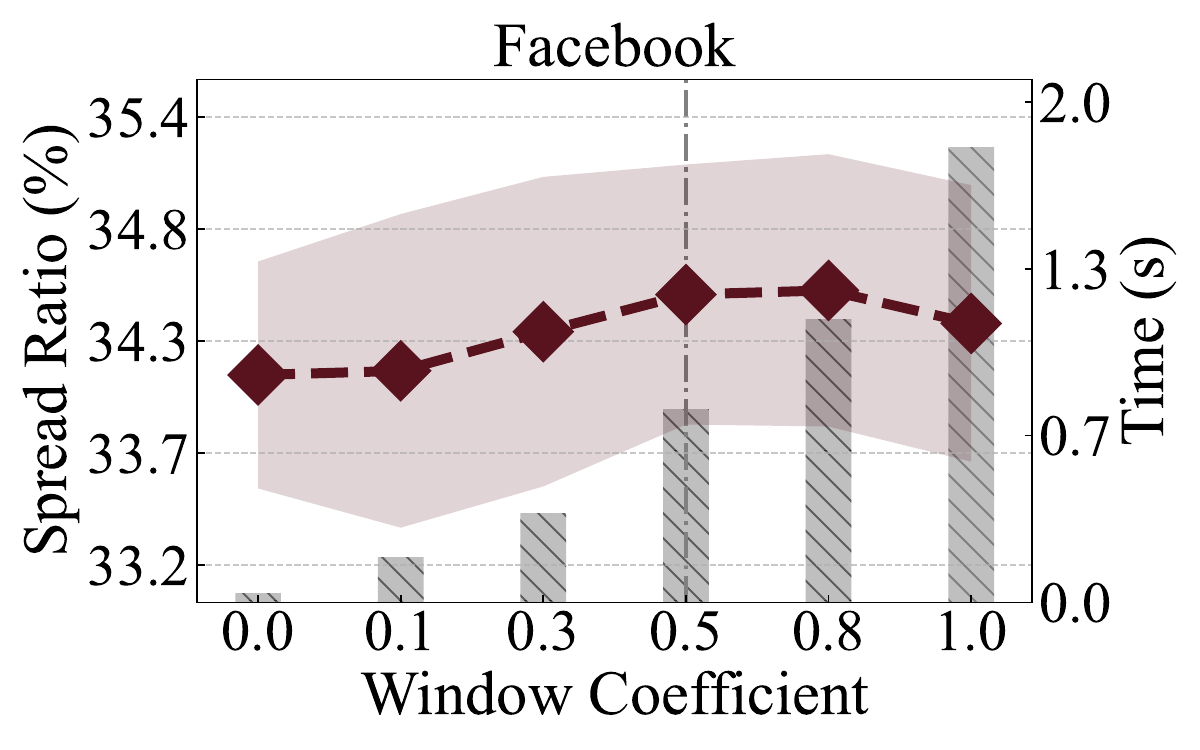}}
    \subfloat{\includegraphics[width=0.38\columnwidth]{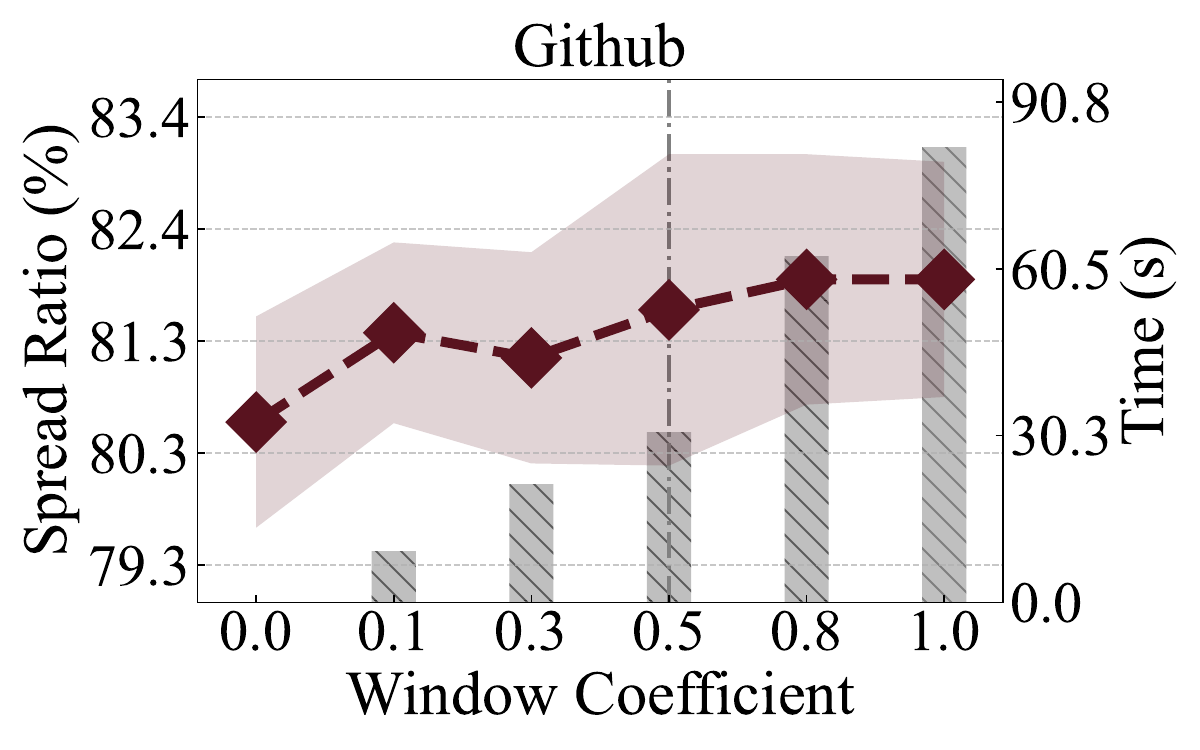}}
    \caption{Results on window coefficient under the WLT model.}
    \label{fig:w_wlt}
\end{figure} 
\textbf{Impact of Window Coefficient}.
To analyze the characteristics of the ASW selection strategy, we investigate how varying the window coefficient $\beta$ affects the performance and efficacy of HIM.
The experimental results on four datasets under the WIC and WLT models are shown in Figures~\ref{fig:w_wic} 
and~\ref{fig:w_wlt}, respectively. 
The line plot shows the spread ratio with standard deviation, while the bar chart shows the selection time.
The seed ratio is $10\%$.
The vertical dashed line indicates the default $\beta$.
We include HIM$_{MD}$ with $\beta=0$ for reference.
Generally, HIM performs better than HIM$_{MD}$ across most $\beta$ values, indicating that it can leverage social relations and inter-user distance to identify seed sets that yield larger influence spread. 
The optimal setting of $\beta$ depends on the dataset scale. 
In general, increasing $\beta$ tends to improve effectiveness but also incurs higher computational cost, leading to a trade-off between effectiveness and efficiency.
In practice, $\beta$ can be flexibly adjusted according to specific requirements.
Overall, the proposed ASW strategy effectively exploits diverse distance information from learned hyperbolic representations to produce high-quality seed sets.

\subsection{Visualization}
\begin{figure}[!ht]
\vspace{-2em}
\centering
    \subfloat{\includegraphics[width=0.40\columnwidth]{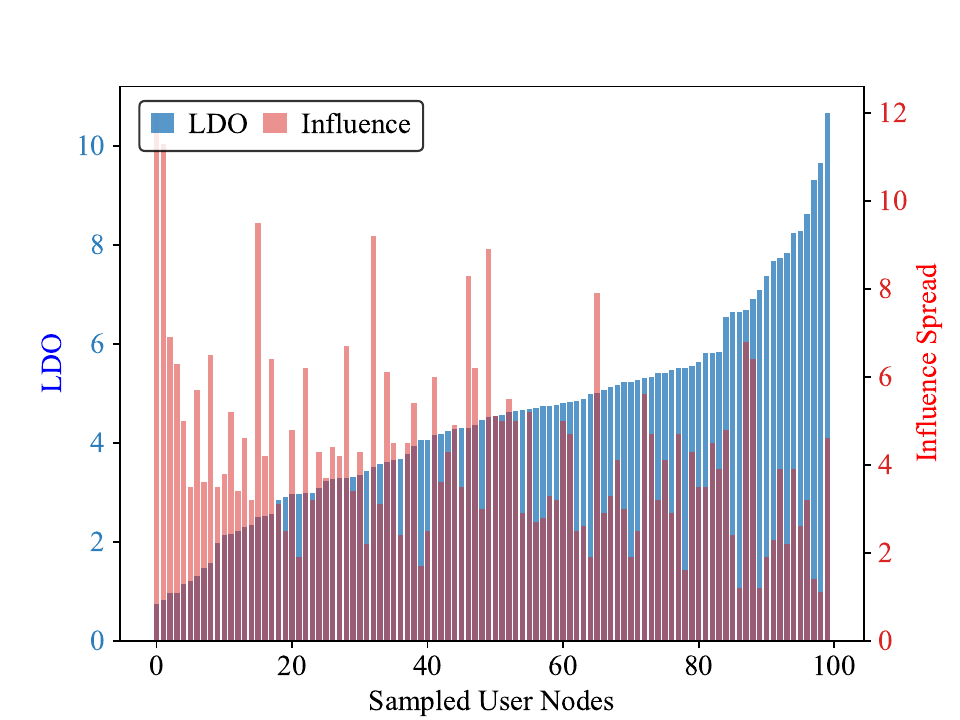}}
    \subfloat{\includegraphics[width=0.40\columnwidth]{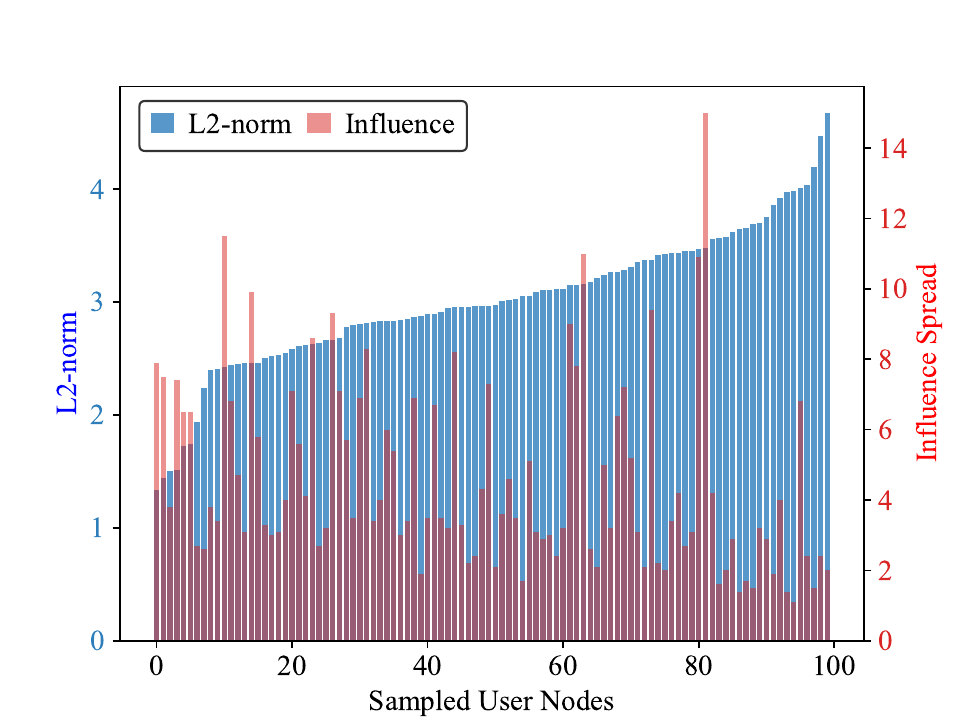}}
    \caption{Distance-influence relation on Power Grid in hyperbolic space (left) and Euclidean space (right).}
    \label{fig:dis_inf}

    \vspace{-1em}
    \subfloat{\includegraphics[width=0.40\columnwidth]{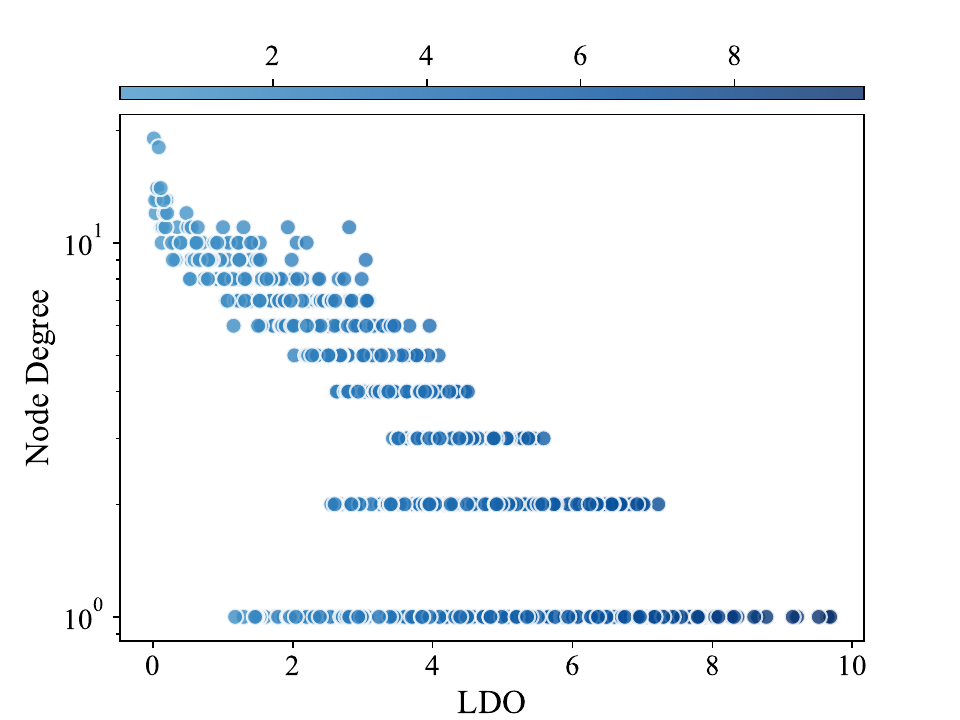}}
    \subfloat{\includegraphics[width=0.40\columnwidth]{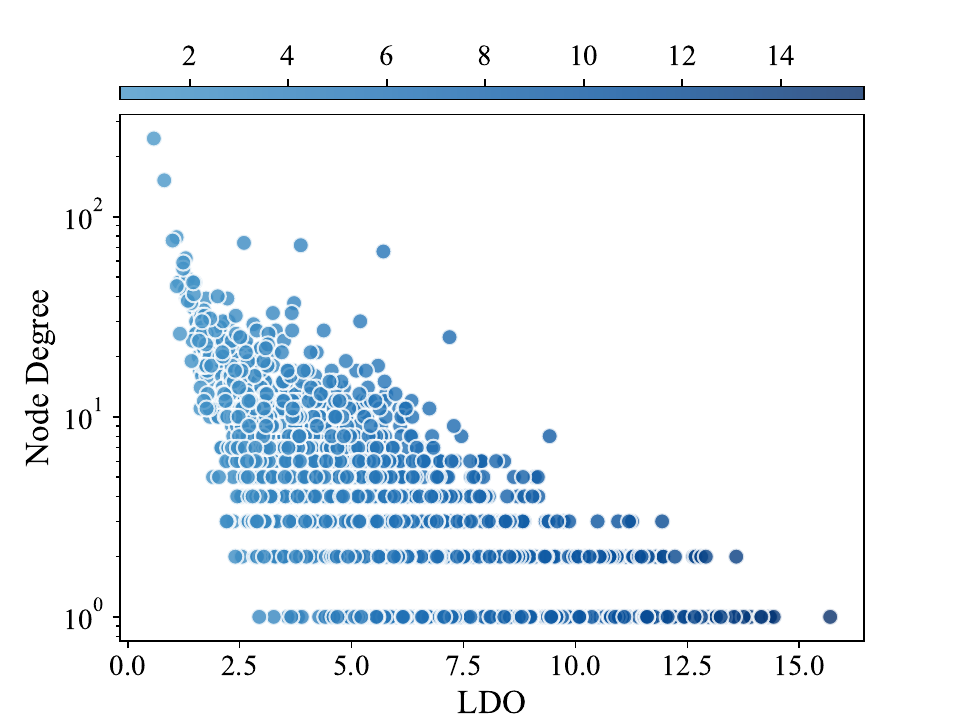}}
    \caption{LDO-degree relations on Power Grid (left) and Cora-ML (right) datasets.}
    \label{fig:ldo_degree}
    \vspace{-1em}
\end{figure}
We investigate the interpretability of our method through visualization on Power Grid and Cora-ML datasets under the IC model. Figure~\ref{fig:dis_inf} shows the relationship between the distances of each user's representations in different spaces, i.e., $LDO$ in hyperbolic space (left) and L2-norm in Euclidean space (right), and their influence strength. For clarity, 100 points were randomly sampled for plotting. The $LDO$ of hyperbolic representations demonstrates a clear inverse relationship with individual user influence, unlike the Euclidean representations. This indicates that hyperbolic representations effectively reflect the magnitude of user influence through $LDO$. 
Moreover, Figure~\ref{fig:ldo_degree} shows the relationship between the hyperbolic distance, i.e., $LDO$, and the degree of nodes on Cora-ML and Power Grid datasets, respectively. 
Intuitively, the degree serves as a rough measure of user influence, assuming that nodes with larger degrees usually tend to be more influential. 
Overall, both figures reveal an inverse relationship between degree and $LDO$, which indicates that the nodes with smaller $LDO$s are more likely to have more connections. In addition, the nodes with the same degree may have different $LDO$s, suggesting that hyperbolic representation can help differentiate nodes with similar degrees. 
Similar trends are also observed in other datasets.
Overall, the results show that our method effectively captures user influence spread patterns with hyperbolic representations.

\tblue{\section{Limitations and Future Work}
Although the theoretical design and experimental results verify the effectiveness of HIM, the current framework may still have several limitations. 
First, our framework assumes that historical propagation instances are available and sufficiently informative to reflect influence dynamics. 
If the observed propagation data are sparse, incomplete, or noisy, the learned hyperbolic representations may become biased, thereby reducing the effectiveness of influence estimation.
This suggests that the quality and informativeness of the available propagation instances remain important for the HIM's performance.
Second, the current representation framework primarily models social structure and propagation relations and does not explicitly incorporate richer factors, such as user attributes, content semantics, or other contextual signals, which may also affect social modeling. 
Nevertheless, we view our method as a fundamental modeling framework in which social structure and propagation relations capture the most essential factors of influence diffusion. 
On this basis, the framework can be further extended to incorporate richer modalities in future work.}

In addition, several promising directions are worth further investigation in future work. 
First, it would be interesting to further investigate how the properties of hyperbolic geometry can be leveraged to design more advanced seed selection strategies.
Second, extending our method to ultra-large-scale social networks with hundreds of millions or even billions of nodes remains an important and challenging direction. 
In addition, the proposed framework may offer a promising direction for supporting more complex and realistic propagation mechanisms, such as invitation-based diffusion~\cite{WWW_zhang_2024_information}, as well as influence propagation on more complex network structures, including hypergraphs~\cite{IPM_xie_2023_efficient}.

\section{Conclusion}
In this work, we propose HIM, a novel diffusion model agnostic method for the IM problem.
The core idea is to estimate users’ influence strength from social data in hyperbolic space to guide seed user selection.
HIM leverages the expressive power of hyperbolic representations to effectively capture complex social influence patterns from the social network and influence propagation instances for the IM problem. 
It is purely data-driven and can operate under various influence propagation scenarios.
We empirically demonstrate the effectiveness of HIM on five real-world networks under various diffusion models with known and unknown propagation parameters. 
The results show that our method consistently achieves strong performance across diverse and complex scenarios. 
Especially when diffusion parameters are unknown, HIM outperforms all baselines across five networks, achieving the highest influence spread across different propagation modes. 
In addition, compared to other graph learning-based methods, HIM exhibits superior scalability, efficiently handling social networks with millions of nodes.
Our work represents a preliminary effort in applying hyperbolic geometry to IM tasks, paving the way for further study in understanding influence propagation using non-Euclidean geometry.

\ifCLASSOPTIONcaptionsoff
  \newpage
\fi



%
\bibliographystyle{IEEEtran}
\bibliography{reference}

%
\vspace{-5em}
\begin{IEEEbiography}[{\includegraphics[width=1in,height=1.1in,clip,keepaspectratio]{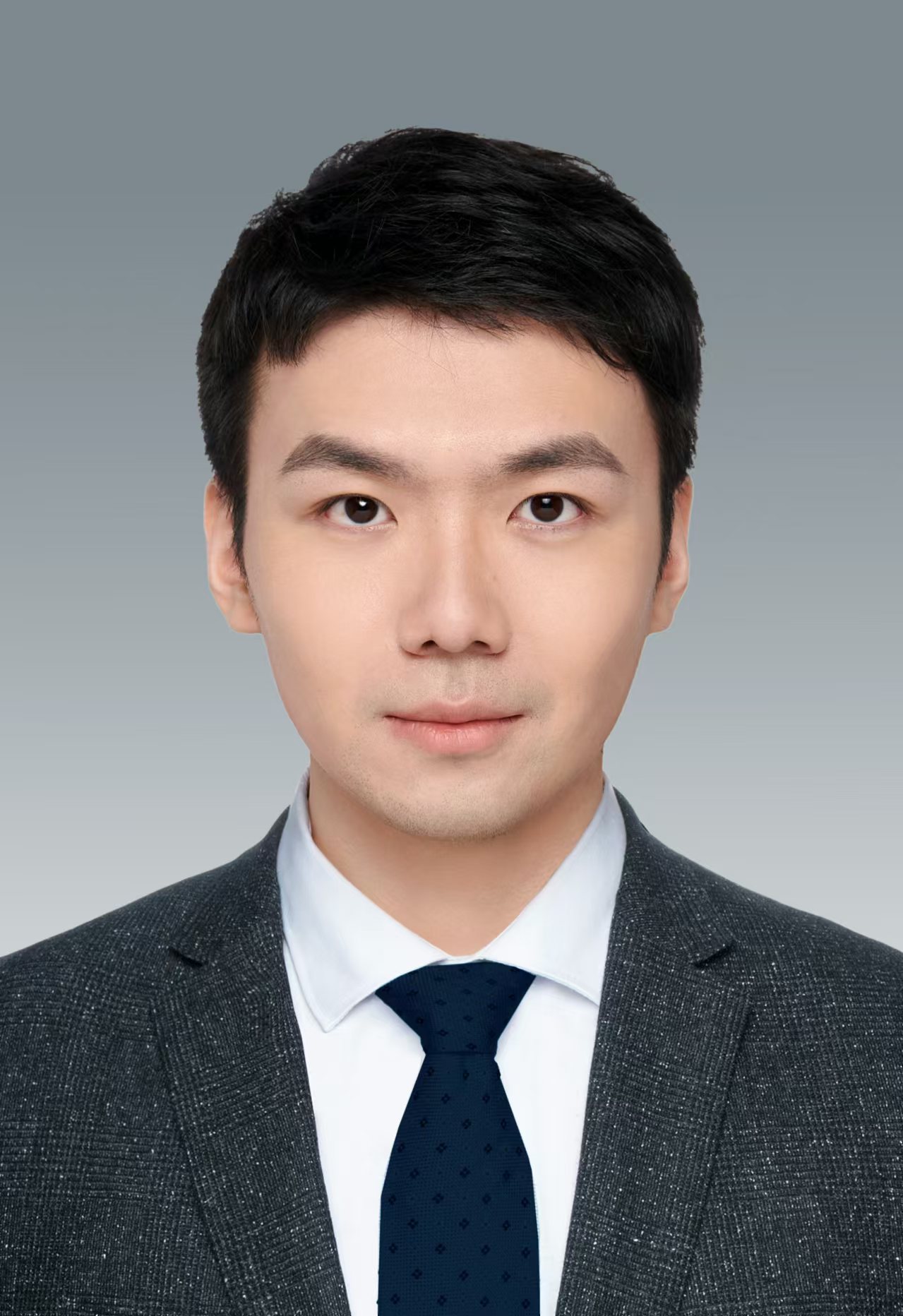}}]
{Hongliang Qiao} 
is currently a Ph.D. student at The Hong Kong Polytechnic University, Hong Kong SAR, China.
His research interests include spatial-temporal data mining and social network analysis.
\end{IEEEbiography}
\vspace{-2.5em}
\begin{IEEEbiography}[{\includegraphics[width=1in,height=1.25in,clip,keepaspectratio]{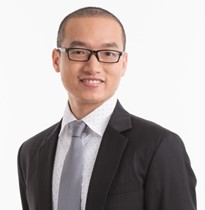}}]
{Shanshan Feng}
received the Ph.D. degree from Nanyang Technological University, Singapore in 2017, and his B.E. degree from the University of Science and Technology of China in 2011. He is currently a Professor at the School of Computer Science, Wuhan University, Wuhan, China. His current research interests include big data analytics, social graph learning, and recommender systems. He has published more than 50 papers in prestigious journals and conferences including IEEE TPAMI, IEEE TKDE, IEEE TNNLS, SIGKDD, SIGIR, ICDE, VLDB, etc. He served as a PC member of SIGIR, SIGKDD, ICDE, ACM MM, NeurIPS, etc.
\end{IEEEbiography}
\vspace{-2.5em}
\begin{IEEEbiography}[{\includegraphics[width=1in,height=1.25in,clip,keepaspectratio]{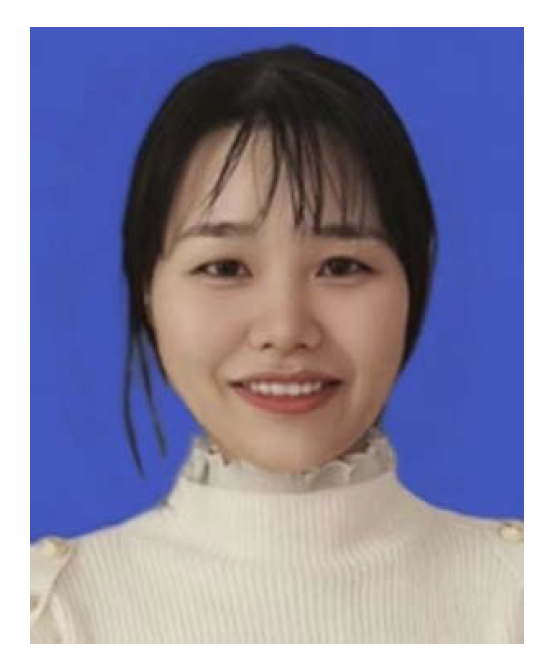}}]
{Min Zhou} received the BS degree in automation from the University of Science and Technology of China, and the PhD degree from Industrial Systems Engineering and Management Department, National University of Singapore, respectively. She is currently a principal research engineer of Huawei Noah’s Ark Lab, Shenzhen, China. Her interests include pattern mining and machine learning, and their applications in sequence and graph data. Her several works related to graph learning and mining were published at top conferences, including KDD, WWW, ICDE, and SIGIR.
\end{IEEEbiography}
\vspace{-2.5em}
\begin{IEEEbiography}[{\includegraphics[width=1in,height=1.1in,clip,keepaspectratio]{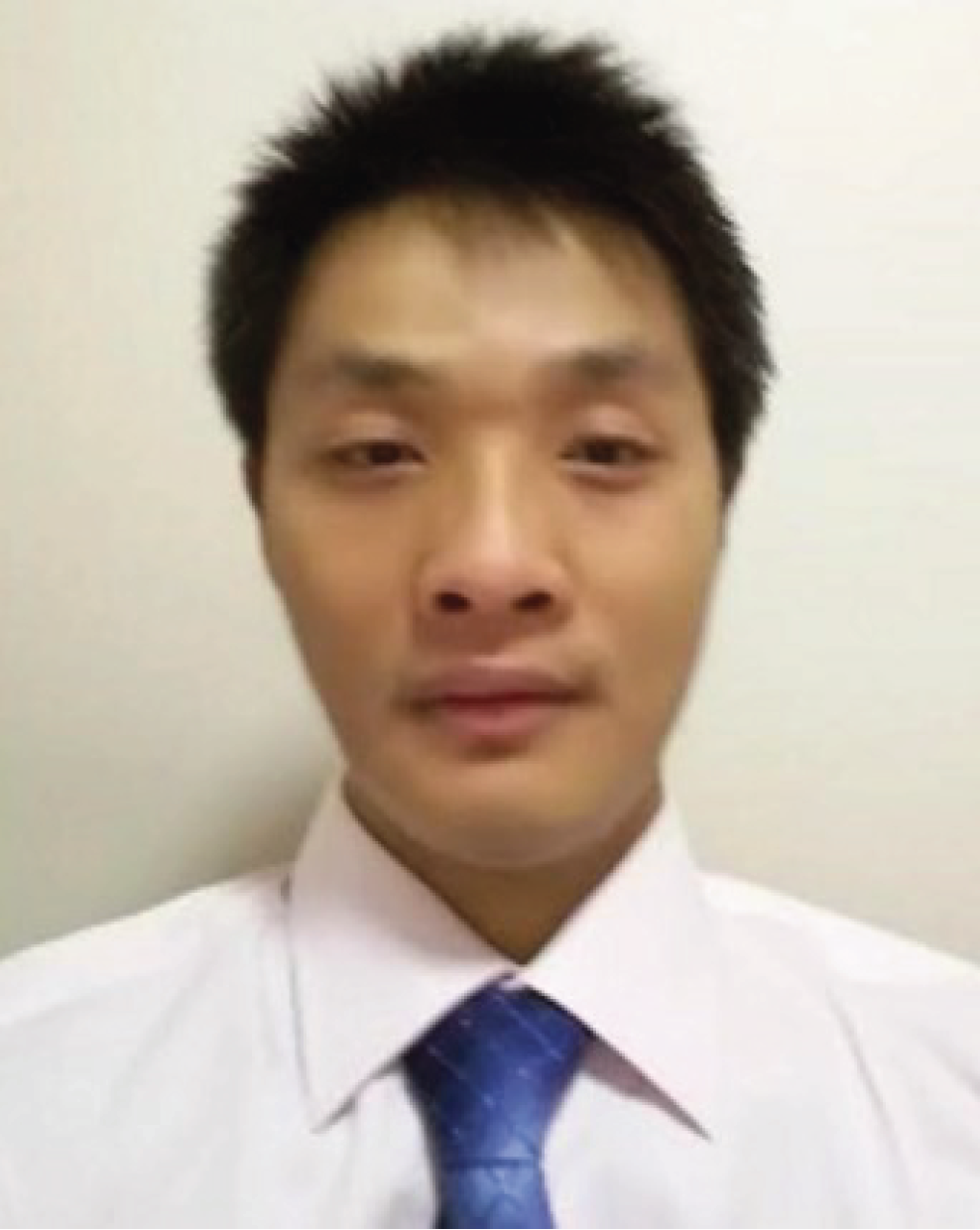}}]
{Xutao Li} is currently a Professor with the School of Computer Science and Technology, Harbin Institute of Technology, Shenzhen, China. He received the Ph.D. and Master degrees in Computer Science from Harbin Institute of Technology in 2013 and 2009, and the Bachelor from Lanzhou University of Technology in 2007. His research interests include data mining, machine learning, graph mining, and social network analysis, especially tensor-based learning and mining algorithms.
\end{IEEEbiography}
\vspace{-2.5em}
\begin{IEEEbiography}[{\includegraphics[width=1in,height=1.1in,clip,keepaspectratio]{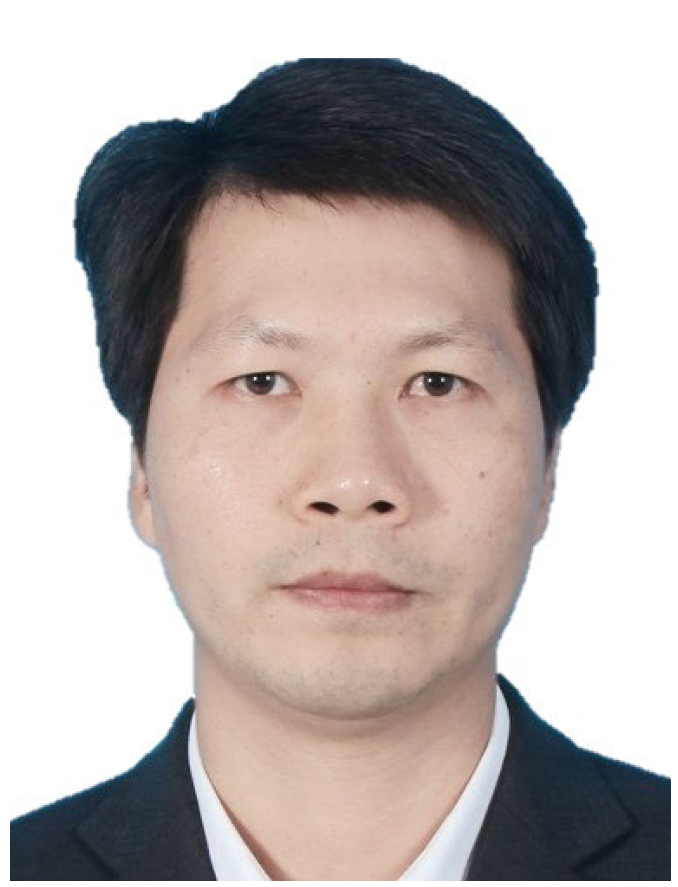}}]
{Yunming Ye} received the Ph.D. degree in computer science from Shanghai Jiao Tong University, Shanghai, China, in 2004.

He is currently a Professor with the School of Computer Science and Technology, Harbin Institute of Technology, Shenzhen, China. His research interests include data mining, text mining, and ensemble learning algorithms.
\end{IEEEbiography}
\vspace{-2.5em}
\begin{IEEEbiography}[{\includegraphics[width=1in,height=1.1in,clip,keepaspectratio]{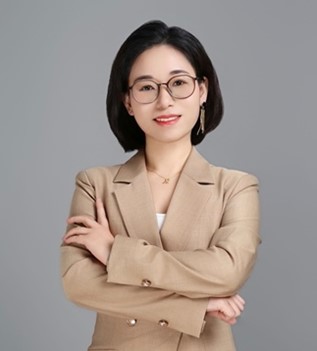}}]
{Fan Li} received her Ph.D. degree from Nanyang Technological University, Singapore, in 2020, and her B.E. degree from Beihang University, Beijing, China, in 2013. She is currently an Assistant Professor in the Department of Aeronautical and Aviation Engineering at The Hong Kong Polytechnic University, Hong Kong SAR, China. Her primary research areas include human-AI teaming, human-AI interaction design, and cognitive ergonomics. 
\end{IEEEbiography}
\vspace{-2.5em}
\begin{IEEEbiography}[{\includegraphics[width=1in,height=1.1in,clip,keepaspectratio]{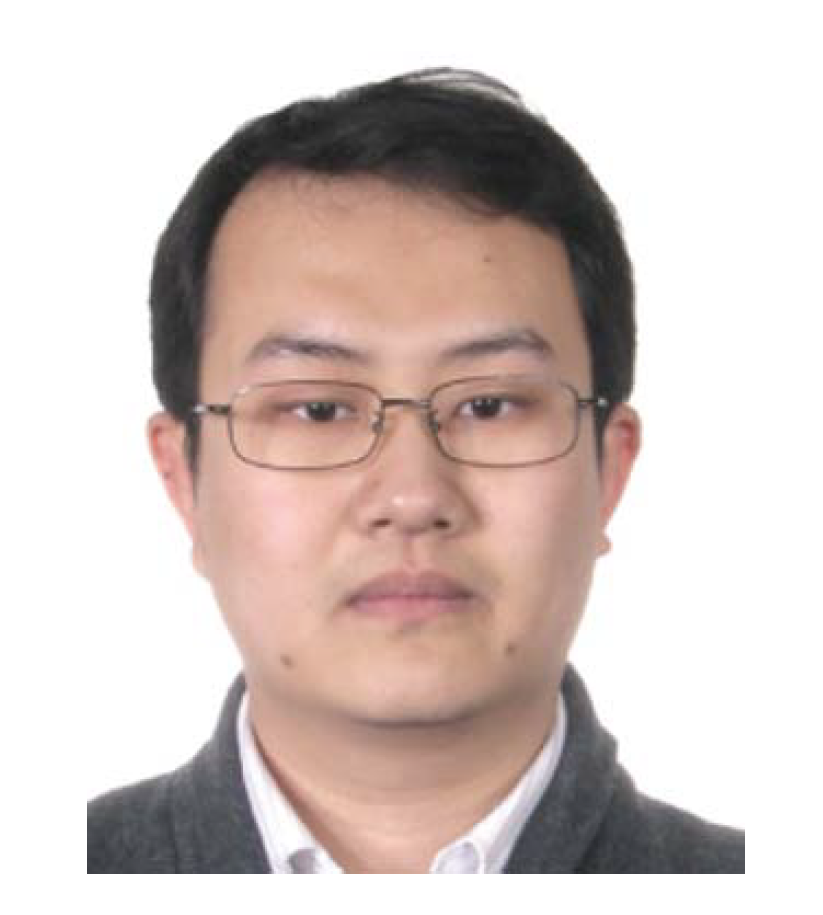}}]
{Shuo Shang} is a professor in computer science with the University of Electronic Science and Technology of China. His research interests include Big Data, big model, and artificial intelligence in general. He is an action editor of Geoinformatica and an associate editor of DSE. He is a regular (senior) TPC Member of many top-ranked conferences including SIGMOD, VLDB, ICDE, KDD, WWW, Neurips, AAAI, and IJCAI.
\end{IEEEbiography}
\vspace{-1.5em}
\begin{IEEEbiography}[{\includegraphics[width=1in,height=1.25in,clip,keepaspectratio]{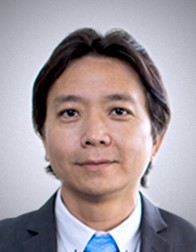}}]
{Yew-soon Ong}
(Fellow, IEEE) received the Ph.D. degree from the University of Southampton, U.K., in 2003. He is the president’s chair professor in Computer Science with Nanyang Technological University (NTU), and holds the position of chief artificial intelligence scientist of A*STAR, Singapore. At NTU, he serves as co-director of the Singtel-NTU Cognitive Artificial Intelligence Joint Lab. His research interest is in artificial and computational intelligence. He is the founding EIC of IEEE Transactions on Emerging Topics in Computational Intelligence and AE of IEEE Transactions on Neural Networks and Learning Systems, IEEE on Transactions on Cybernetics, IEEE Transactions on Artificial Intelligence and others. He has received several IEEE outstanding paper awards and was listed as a Thomson Reuters highly cited Researcher and among the World’s Most Influential Scientific Minds.
\end{IEEEbiography}




\end{document}